\documentclass[cite]{elsarticle}

\journal{Nuclear Physics A}
\usepackage{graphicx,epsfig,wrapfig,color}
\usepackage{amsmath}
\usepackage{amssymb}
\newcommand{\be}{\begin{equation}}
\newcommand{\ee}{\end{equation}}
\newcommand{\ba}{\begin{eqnarray}}
\newcommand{\ea}{\end{eqnarray}}
\newcommand{\di}{\!{\rm d}}
\newcommand{\la}{\langle}
\newcommand{\ra}{\rangle}
\newcommand{\eps}{\varepsilon}
\newcommand{\teps}{\tilde{\varepsilon}}
\newcommand{\rphi}{\tilde{\varphi}}
\begin{document}
\begin{frontmatter}
\title{The energy-momentum tensor and D-term of Q-clouds}	

\author[UConn]{Michael Cantara}
\author[Yale]{Manuel Mai}
\author[UConn]{Peter Schweitzer}

\address[UConn]{Department of Physics, University of Connecticut, 
		Storrs, CT 06269-3046, U.S.A.}
\address[Yale]{	Department of Physics Yale University,
   		New Haven, CT 06511-8499, U.S.A.}

\begin{abstract}
The $D$-term is, like mass and spin, a fundamental property related 
to the energy-momentum~tensor. Yet it is not known experimentally for any 
particle. In all theoretical studies so far the $D$-terms of various 
particles were found negative. Early works gave rise to the assumption 
the negative sign could be related to stability. The emerging question 
is whether it is possible to find  a field-theoretical 
system with a positive $D$-term. 
To shed some light on this question we investigate 
$Q$-clouds, an extreme parametric limit in the $Q$-ball system.
$Q$-clouds are classically unstable solutions which delocalize, spread out 
over all space forming an infinitely dilute gas of free quanta, and are
even energetically unstable against tunneling to plane waves.
In short, these extremely unstable field configurations provide 
an ideal candidate system for our purposes.
By studying the energy-momentum tensor we show that at any stage
of the $Q$-cloud limit one deals with perfectly well-defined and, 
when viewed in appropriately scaled coordinates, non-dissipating 
non-topological solitonic solutions. We investigate in detail their 
properties, and find new physical interpretations by observing that 
$Q$-clouds resemble BPS Skyrmions in certain aspects, and correspond to 
universal non-perturbative solutions in  (complex) $|\Phi|^4$ theory. 
In particular, we show that also $Q$-cloud solutions have negative
$D$-terms. Our findings do not prove that $D$-terms must always be 
negative. But they indicate that it is unlikely to realize a positive 
$D$-term in a consistent physical system.
\end{abstract}

\begin{keyword}
energy momentum tensor, $Q$-ball, soliton, stability, $D$-term
\end{keyword}

\end{frontmatter}


\section{Introduction}
\label{Sec-1:introduction}

The $D$-term \cite{Polyakov:1999gs} is a particle property as fundamental
as mass or spin, yet not known for any particle. It is defined 
through form factors of the energy momentum tensor \cite{Pagels}.
As the electric form factor provides information on the charge 
distribution \cite{Sachs}, so does the form factor associated with the 
$D$-term give insights into the distribution of internal forces inside
a particle 
(and the interpretation is subject to the same type of limitations)
\cite{Polyakov:2002yz}.

Information on the $D$-term can be accessed in hard exclusive 
reactions \cite{Muller:1998fv,Ji:1998pc,Teryaev:2001qm}.
Theoretical studies dedicated to the $D$-term include 
soft pion theorems, chiral perturbation theory, lattice QCD, soliton 
models, nuclear models, bag and spectator models, and dispersion techniques
\cite{Polyakov:1999gs,Polyakov:2002yz,
Ji:1997gm,Petrov:1998kf,
Goeke:2007fp,Goeke:2007fq,Cebulla:2007ei,
Kim:2012ts,Mueller:2011bk,Pasquini:2014vua,
Donoghue:1991qv,Megias:2004uj,
Mathur:1999uf,Liuti:2005gi,Gabdrakhmanov:2012aa,Mai:2012yc,Mai:2012cx}.
Remarkably, in all theoretical studies so far the $D$-terms
(of pions, nucleons, nuclei, photons, $Q$-balls) were found 
to be negative. 

In early works a connection was suspected between the sign of the 
$D$-term and the stability of a particle \cite{Goeke:2007fp}. However, 
insightful studies in the $Q$-ball system \cite{Mai:2012yc,Mai:2012cx} 
revealed that meta-stable, unstable and excited solutions (which all still 
correspond to local minima of the action) have also negative $D$-terms.

An emerging question that motivates our study is whether the D-term 
can be positive in a physical system. To address this question, we 
will use the $Q$-ball system as a theoretical laboratory once more, and 
investigate a particular limit in which unstable solutions ``dissociate'' 
into a ``cloud'' of free quanta. 

$Q$-balls are non-topological solitons in theories with global symmetries 
\cite{Friedberg:1976me,Coleman:1985ki,Safian:1987pr}. They may have been 
created under the conditions of the early universe, were discussed 
as dark matter candidates, and have applications in astrophysics,
cosmology, and particle physics
\cite{Cohen:1986ct,Alford:1987vs,Lee:1991ax,Kusenko:1997ad,
Kasuya:1999wu,Multamaki:1999an,Clark:2005zc,Clougherty:2005qg,
Schmid:2007dm,Verbin:2007fa,Volkov:2002aj,Gleiser:2005iq,Gani:2007bx,
Sakai:2007ft,Tsumagari:2008bv,Bowcock:2008dn,Arodz:2008jk,Gabadadze:2008sq,
Campanelli:2009su,Krylov:2013qe,Nugaev:2013poa}.

Studies of the parametric limit in which unstable $Q$-balls 
delocalize, spread out over all space, and eventually form 
an infinitely dilute system of free quanta, date back to 
\cite{Alford:1987vs} where the interpretation as a ``$Q$-cloud'' 
was given. 

$Q$-clouds are extreme, unstable field configurations. 
In contrast to unstable $Q$-balls or excited $Q$-ball states
shown to have negative $D$-terms \cite{Mai:2012yc,Mai:2012cx} 
and decaying in several 
lighter stable $Q$-balls of the same total charge, $Q$-clouds
are even energetically unstable against the tunneling to plane waves.

$Q$-ball solutions have been subject to modest interest in literature so far.  
Interesting recent developments are the demonstration of the existence 
of $Q$-clouds around Kerr black holes leading to the discovery
of a new family of hairy black holes \cite{Herdeiro:2014pka}, and
a possible connection to sphalerons \cite{Nugaev:2015rna}. Noteworthy 
is also the possibility to realize experimentally $Q$-cloud type 
configurations in ultra-cold Bose gases~\cite{Khlebnikov:1999qy}. 
The modest attention $Q$-balls have received so far 
is perhaps related to their extreme instability. However, precisely this
property makes $Q$-balls an ideal theoretical testing ground for the 
purposes of our work. 
Could such an extreme and unstable 
system exhibit a positive $D$-term?

In this work, we consider a scalar theory with U(1) symmetry 
and potential~$V$ which admits solitons of the form 
$\Phi(\vec{r},t)=\phi(r)\,e^{i\omega t}$ 
for $\omega_{\rm min}<\omega<\omega_{\rm max}$ with limiting 
frequencies fixed in terms of the potential.
For $\omega$ approaching $\omega_{\rm min}$  one deals with stable 
$Q$-balls which are characterized by a constant charge density 
\cite{Coleman:1985ki} and share many characteristics of fluid 
drops \cite{Mai:2012yc}. 

In the opposite limit $\omega\to\omega_{\rm max}$
the solutions are unstable. Their mass $M$ approaches from above 
$m Q$, where $Q$ is the charge of the solutions and $m$ denotes the 
mass of the elementary quanta \cite{Alford:1987vs}.
Some properties of the solutions as $\omega$ approaches $\omega_{\rm max}$
were studied in \cite{Mai:2012yc}. But many questions remain open.

What are the behavior and the properties of the solutions as
$\omega\to\omega_{\rm max}$? Does a well-defined limiting solution exist?
And, to iterate the central question that motivates this work:
considering that the dissociation of $Q$-matter into a $Q$-cloud
constitutes a genuine instability, could one encounter in this system a 
positive $D$-term? 
The purpose of this work is to address these questions.

The outline is as follows. 
In Sec.~\ref{Sec-2:EMT-and-Qballs} we review $Q$-balls and their properties.
In Sec.~\ref{Sec-3:towards-the-limit} we use the Newtonian interpretation
of the equations of motion to establish qualitative expectations in the limit 
$\omega\to\omega_{\rm max}$. In Sec.~\ref{Sec-4:rescaling} we demonstrate the 
existence of a limiting solution in terms of appropriately defined scaled 
coordinates and fields.
The Secs.~\ref{Sec-5:densities} and \ref{Sec-6:global-properties}
are dedicated to the study of local and global properties as
$\omega$ approaches $\omega_{\rm max}$.
In Sec.~\ref{Sec-7:interpretation} we discuss the properties
of the limiting solution, and interpret the results.
Conclusions are presented in Sec.~\ref{Sec-8:conclusions}.
Technical details are addressed in Appendices.

\section{\boldmath  The $Q$-ball system}
\label{Sec-2:EMT-and-Qballs}

We study the complex scalar field theory defined in terms of the
Lagrangian \cite{Coleman:1985ki} 
with an effective  (non-renormalizable) sixtic potential
\ba\label{Eq:Lagrangian}
	&&{\cal L} = \frac12\,(\partial_\mu\Phi^\ast)(\partial^\mu\Phi) - V,\\
\label{Eq:potential}
   	&&V = A\,(\Phi^\ast\Phi)-B\,(\Phi^\ast\Phi)^2+C\,(\Phi^\ast\Phi)^3\,.
\ea
The positive constants $A$, $B$, $C$ are such that 
\be\label{Eq:xi}
	0 < \xi < 1 \, , \;\;\; 
	\xi = \frac{B^2}{4AC} \equiv \frac{1}{2\alpha}\;,
\ee
which guarantees that $V\ge0$ $\forall\;\Phi$, and $V>0$ if $\Phi\neq 0$.
We define in (\ref{Eq:xi}) the constant $\alpha$ for later convenience.

Non-topological solitons exist due to the global U(1) symmetry
($\Phi\to\Phi\,e^{i\eta}$, $\Phi^\ast\to\Phi^\ast e^{-i\eta}$, 
$\eta\in\mathbb{R}$) of ${\cal L}$. 
In the soliton rest frame the solutions are given by
\be\label{Eq:ansatz}
   \Phi(t,\vec{x}) = \exp(i\omega t)\,\phi(r)\;,\;\;\;r=|\vec{x}\,|\:,
\ee
where $\omega>0$ can be chosen without loss of generality, 
and $\phi(r)$ satisfies the following equation and boundary conditions 
(the primes denote differentiations with respect to the respective arguments)
\ba
&&	\phi^{\prime\prime}(r)+\frac2r\;\phi^\prime(r)+\omega^2\phi(r)
      	- V^\prime(\phi) = 0\:,  \label{Eq:eom}\\
&&  	\phi(0) \equiv \phi_0 = {\rm const},   \;\;
    	\phi^\prime(0) = 0,   \;\nonumber\\
&&   	\phi(r)\to 0\;\;\mbox{for}\;\;r\to\infty\;.
	\label{Eq:boundary-conditions}\ea

\begin{figure}[b!]
\centering
\includegraphics[width=8cm]{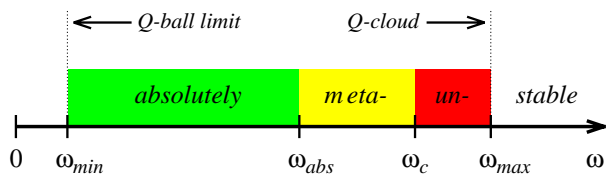}
\caption{\label{FIG-01:omega-range}
	The solitons of the theory 
	(\ref{Eq:Lagrangian}) are absolutely stable 
	for  $\omega_{\rm min}<\omega<\omega_{\rm abs}$, classically
	stable for $\omega_{\rm abs}<\omega<\omega_c$, and unstable
	for $\omega_c<\omega<\omega_{\rm max}$. 
	The arrows indicate the $Q$-ball limit \cite{Coleman:1985ki} 
	and the $Q$-cloud limit \cite{Alford:1987vs}.
	The numerical values of the $\omega_i$ depend on 
	the parameters of the theory in Eq.~(\ref{Eq:potential}).}
\end{figure}

Finite energy solutions exist for $\omega$ in the range
\cite{Coleman:1985ki} 
\ba\label{Eq:condition-for-existence}
   &&  	\omega_{\rm min} < \omega < \omega_{\rm max} \;,\phantom{\biggl|}\\
   &&   \omega_{\rm min}^2 = \min\limits_\phi \biggl[\frac{2\,V(\phi)}{\phi^2}
        \biggr] = 2A(1-\xi)>0\;, \nonumber\\
   &&   \omega_{\rm max}^2 = V^{\prime\prime}(\phi)\biggl|_{\phi=0}=2A\;.
    	\nonumber\ea 
Notice that $\omega_{\rm max}$ defines the mass $m$ of the elementary
quanta of the field $\Phi$ as 
\be\label{Eq:mass-elementary-quantum}
	m=\omega_{\rm max} = \sqrt{2A}\,.
\ee
The solutions can be classified as sketched in Fig.~\ref{FIG-01:omega-range}.

For $\omega_{\rm min}<\omega<\omega_{\rm abs}$ one encounters stable solutions 
satisfying the absolute stability condition $M < m\,Q$ \cite{Lee:1991ax}, 
where $M$ and $Q$ denote the soliton mass and charge. The term ``$Q$-ball'' 
was originally coined to denote solutions in the limit of $\omega$ 
approaching $\omega_{\rm min}$ \cite{Coleman:1985ki}.

For $\omega_{\rm abs}<\omega<\omega_c$ the  solutions are meta-stable,
i.e.\ stable with respect to small fluctuations 
\cite{Friedberg:1976me,Lee:1991ax}.
They satisfy a weaker ``classical stability condition'' which can be 
expressed, for instance, as $Q^{\,\prime}(\omega) \le 0$. At the critical 
frequency $\omega_c$ the charge becomes minimal.

For $\omega_c<\omega<\omega_{\rm max}$ the unstable solutions can
decay into stable $Q$-balls with the same charge but a lower total mass. 
In the limit $\omega\to\omega_{\rm max}$ one then deals with $Q$-clouds
\cite{Alford:1987vs}.

To conduct our study we will investigate properties related to the 
charge and energy-momentum tensor $T^{\mu\nu}$ (EMT), which are 
introduced in the following.

The conserved charge due to the U(1) symmetry is
\be\label{Eq:charge}
   Q = \int\di^3r\;\rho_{\rm ch}(r)\;,\;\;\;
   \rho_{\rm ch}(r) = \omega\;\phi(r)^2\,.
\ee
The canonical EMT of the theory (\ref{Eq:Lagrangian}) is symmetric, 
and static for the solutions (\ref{Eq:ansatz}) \cite{Mai:2012yc}. 
The energy density
\be\label{Eq:T00}
   T_{00}(r) = \frac12\,\omega^2\phi(r)^2+\frac12\,\phi^\prime(r)^2+V(\phi)\;,
\ee
defines the mass $M=\int\di^3r\,T_{00}$. One finds $T_{0k}=0$, 
i.e.\ our solutions have spin zero
	(see \cite{Volkov:2002aj} for a
	discussion of spinning solutions).
The spatial components 
\be\label{Eq:Tik}
   T_{ij} = \biggl(\frac{r_ir_j}{r^2}-\frac13\,\delta_{ij}\biggr) s(r)
   + \delta_{ij}\,p(r)
\ee
define the stress tensor where $s(r)$ and $p(r)$ denote 
the distributions of shear forces and pressure given by
\ba
    s(r) &=& \phi^\prime(r)^2 \label{Eq:shear}\\
    p(r) &=& \frac12\,\omega^2\phi(r)^2-\,\frac16\,\phi^\prime(r)^2 -V(\phi)
    \label{Eq:pressure}\;.
\ea
The conservation of the EMT dictates that $s(r)$ and $p(r)$ are 
connected by the relation \cite{Polyakov:2002yz}
\be\label{Eq:diff-eq-s-p}
 	\frac{2}{r}\,s(r) + \frac{2}{3}\,s^\prime(r) + p^\prime(r) = 0\;,
\ee
and $p(r)$ must obey \cite{Goeke:2007fp} the von Laue condition
\cite{von-Laue}, a necessary condition for stability,
\be\label{Eq:stability-condition}
      \int_0^\infty \di r\;r^2p(r)=0\;.
\ee

The constant $d_1$, to which we refer here as the $D$-term, 
completes the information content of the EMT and is given 
in terms of $s(r)$ or $p(r)$ as \cite{Polyakov:2002yz}
\ba
    d_1 &=& \phantom{-}\; 5\pi\,M
	\int_0^\infty\di r\;r^4\,p(r) \label{Eq:def-d1-pressure}\;, \\
    	&=& -\,\frac{4\pi}{3}\,M
	\int_0^\infty\di r\;r^4\,s(r) \label{Eq:def-d1-shear}\;.
\ea

Other quantities of interest are ``surface tension'' $\gamma$, 
mean square radius $\la r^2_s\ra$ of the shear forces, 
\ba\label{Eq:def-gamma-rs}
     \gamma = \int_0^\infty\,\di r\;s(r) \; , \;\;\;
     \la r^2_s\ra=\frac{1}{\gamma}\,\int_0^\infty\,\di r\;r^2 s(r)\,,
\ea
and mean square radii of the energy and charge densities
\ba\label{Eq:def-mean-square-radii-E-M}
     \la r^2_E\ra=\frac{\int\di^3r\;r^2\,T_{00}(r)}
                       {\int\di^3r\,     T_{00}(r)} \,,\;\;
     \la r^2_Q\ra=\frac{\int\di^3r\;r^2\,\rho_{\rm ch}(r)}
                       {\int\di^3r\,     \rho_{\rm ch}(r)}\,.\;\;
\ea From Eqs.~(\ref{Eq:eom},~\ref{Eq:boundary-conditions}) one
deduces that the $\phi(r)$ behave as
\ba
\label{Eq:asymp-small}
    \phi(r) &=&
    \phi_0 + \biggl(V^\prime(\phi_0)-\omega^2\phi_0\biggr)\,\frac{r^2}{6}
    + \dots  \;\;\mbox{small $r$},\;\;\;\;\\
    \phi(r)&=&\!\frac{c_\infty}{r}\;
    \exp\biggl(-r\sqrt{\omega_{\rm max}^2-\omega^2}\biggr)\,
    + \dots  \;\mbox{large $r$},
\label{Eq:asymp-large}
\ea
where the dots indicate subleading terms, and $\phi_0$ and $c_\infty$ 
follow from solving the boundary value problem 
(\ref{Eq:eom},~\ref{Eq:boundary-conditions}).

When presenting numerical results we use the same parameters
and $\omega$-range as in Ref.~\cite{Mai:2012yc},
\ba	\label{Eq:parameters}
&	A=1.1\,,\;\;\;\;
	B=2.0\,,\;\;\;\;
	C=1.0\,, & \\
	\label{Eq:omega-range}
&	\omega_{\rm min}^2=0.2 \, , \;\;\;
	\omega_{\rm max}^2=2.2 \, . &
\ea
With these parameters the other frequencies in Fig.~\ref{FIG-01:omega-range}
take the values $\omega_{\rm abs}^2\approx 1.55$ and 
$\omega^2_{\rm c}\approx 1.9$ \cite{Mai:2012yc}.

\section{Towards the $Q$-cloud limit}
\label{Sec-3:towards-the-limit}

\begin{figure}[b!]
\centering
\includegraphics[height=4.5cm]{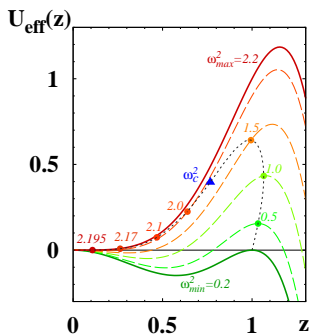}
\caption{\label{FIG-02:eff-pot}
	$U_{\rm eff}(z)=\frac12\,\omega^2z^2-V(z)$ as function of $z$ for the 
	limiting values $\omega_{\rm min}^2=0.2$ and $\omega_{\rm max}^2=2.2$ 
	(solid lines), and for the selected values 
	$\omega^2=0.5,\,1.0,\,1.5,\,2.0$ (dashed lines).
	The circles show the initial values $z_0$ for each $\omega^2$, 
	which lie on a continuous (dotted) curve \cite{Mai:2012yc}. 
	The starting points $z_0$ for $\omega_c^2\simeq 1.9$ and
	 $\omega^2=2.1,\,2.17,\,2.195$ are indicated,
	without plotting $U_{\rm eff}(z)$ for these values.}

\end{figure}

We first establish qualitative expectations in the limit 
$\omega\to\omega_{\rm max}$. 
For that we explore the analogy of (\ref{Eq:eom}) to a Newtonian equation 
for the 1D motion of a unit mass particle described by the coordinate 
$z(t)$, which is subject to a time- and velocity-dependent frictional 
force $F_{\rm fric}(\mbox{\it\.z},t)$ in an effective potential 
$U_{\rm eff}(z)$ defined as
\ba
\label{Eq:Newtonian-eom}
&& \mbox{\it\"z}(t)=F_{\rm fric}(\mbox{\it\.z},t)-\nabla U_{\rm eff}(z) \\
&& F_{\rm fric}(\mbox{\it\.z},t)=-\frac2t\,\mbox{\it\. z}(t) \,,\;\;\;
U_{\rm eff}(z)=\frac12\omega^2\,z^2-V(z) \,,\nonumber
\ea
where we identify $r\leftrightarrow t$ and $\phi(r)\leftrightarrow z(t)$.
At $t=0$ the particle starts with zero velocity (analog to $\phi^\prime(0)=0$)
from the position $z_0$ (analog to $\phi_0$) chosen such that the particle 
will ``slide down'' in the potential $U_{\rm eff}(z)$ shown in 
Fig.~\ref{FIG-02:eff-pot} and, after infinitely long time, stop 
at the origin (analog to $\phi(r)\to 0$ for $r\to\infty$).

This very fruitful analogy was used in \cite{Coleman:1985ki} 
to illustrate the existence of the solutions.
Only if $\omega>\omega_{\rm min}$ can solutions exist, since then there is 
a starting point $z_0 > 0$ where the particle has a potential energy 
$U_{\rm eff}(z_0)>0$ so it can overcome the friction and make it to the origin.
As long as $\omega<\omega_{\rm max}$ the potential  $U_{\rm eff}(z)$ will 
also dip below zero somewhere between $z_0$ and the origin, which is
necessary to allow the particle to dissipate its initial energy 
before arriving at the origin.

When $\omega$ is close to $\omega_{\rm min}$ one deals with ``$Q$-balls,'' 
i.e.\ solutions which exhibit extended plateaus in the inner 
region with nearly constant charge density \cite{Coleman:1985ki}
and resemble fluid drops in many aspects \cite{Mai:2012yc}. 

Here we are interested in the opposite limit $\omega\to\omega_{\rm max}$. 
In this regime the region of $z$ where $U_{\rm eff}(z)<0$ shrinks, causing
the initial positions $z_0$ (from which the particle has 
to be released to arrive at the origin with zero velocity) 
to decrease rapidly, see Fig.~\ref{FIG-02:eff-pot}.

In the language of fields, this means the magnitudes of $\phi(r)$ 
and hence the charge and energy densities decrease. At the same time, 
the fields decay at large $r$ more and more slowly according to 
(\ref{Eq:asymp-large}), implying that the spatial extension of the 
field configurations grows. The decrease of charge and energy
densities is overall overwhelmed by the growth of the spatial extension
of the solutions. As a consequence the total charge and mass diverge. 

The patterns how charge, mass and size of the solutions diverge
were studied numerically in \cite{Mai:2012yc}. To make definite
statements, it is convenient to define
\be\label{Eq:epsilon}
    \eps=\sqrt{\omega_{\rm max}^2-\omega^2}>0 \,.
\ee
It was found that $M$, $Q$, and mean radii diverge like $1/\eps$, and 
$d_1\propto 1/\eps^2$, while the surface tension $\gamma\propto\eps^3$
\cite{Mai:2012yc}.

As  $\omega^2$ reaches $\omega_{\rm max}^2$ the effective
potential $U_{\rm eff}(z)$ never dips below zero. At first glance,
the only viable solution seems to be when the particle is placed
at $z_0=0$ which corresponds to a trivial vacuum solution $\phi(r)=0$
\;$\forall\,r$.
There is, however, also a non-trivial ``critical'' solution,
which is best seen in terms of adequately scaled units.

\section{Rescaling \& existence of limit}
\label{Sec-4:rescaling}

To study the $Q$-cloud limit rigorously we introduce 
dimensionless coordinates $\vec{x}\to \eps\,\vec{r}$ with 
$x=|\vec{x}\,|=\eps\, r$, and define dimensionless 
rescaled fields $\rphi(x)$ as 
\be\label{Eq:rescaling}
	\phi(r) = \eps\;\frac{\rphi(x)}{\sqrt{B}}\; ,
	\;\;\; x = \eps\, r\,.
\ee
For later convenience we introduce $1/\sqrt{B}$ in  (\ref{Eq:rescaling}) 
with the positive, dimensionless parameter $B$ of the potential 
(\ref{Eq:potential}). We also define the dimensionless quantity 
\be\label{Eq:teps}
	\teps^2 = \frac{\eps^2}{m^2} \equiv \frac{\eps^2}{2A}\;,\;\;\;
	0 < \teps^2 < \xi\,,
\ee
whose range follows from (\ref{Eq:condition-for-existence}).
In terms of the rescaled coordinates and fields, the equation 
of motion becomes 
\begin{align}
&	\rphi^{\prime\prime}(x)+\frac2x\;\rphi^\prime(x)-\rphi(x) +4\,
	\rphi^3(x) - 6\,\alpha\,\teps^2\rphi^5(x) = 0 \, ,
	\;\;\;\; \nonumber\\
&	\rphi(0) = {\rm const}, \;\;
	\rphi^\prime(0) = 0, \;\;
	\rphi(x) \to 0 \;\; \mbox{as} \;\; x\to\infty, 
	\label{Eq:eom-resc} 
\end{align}
with the parameter $\alpha = 2AC/B^2$ as defined in Eq.~(\ref{Eq:xi}).

\begin{figure}[b!]
\centering
\includegraphics[width=5.8cm]{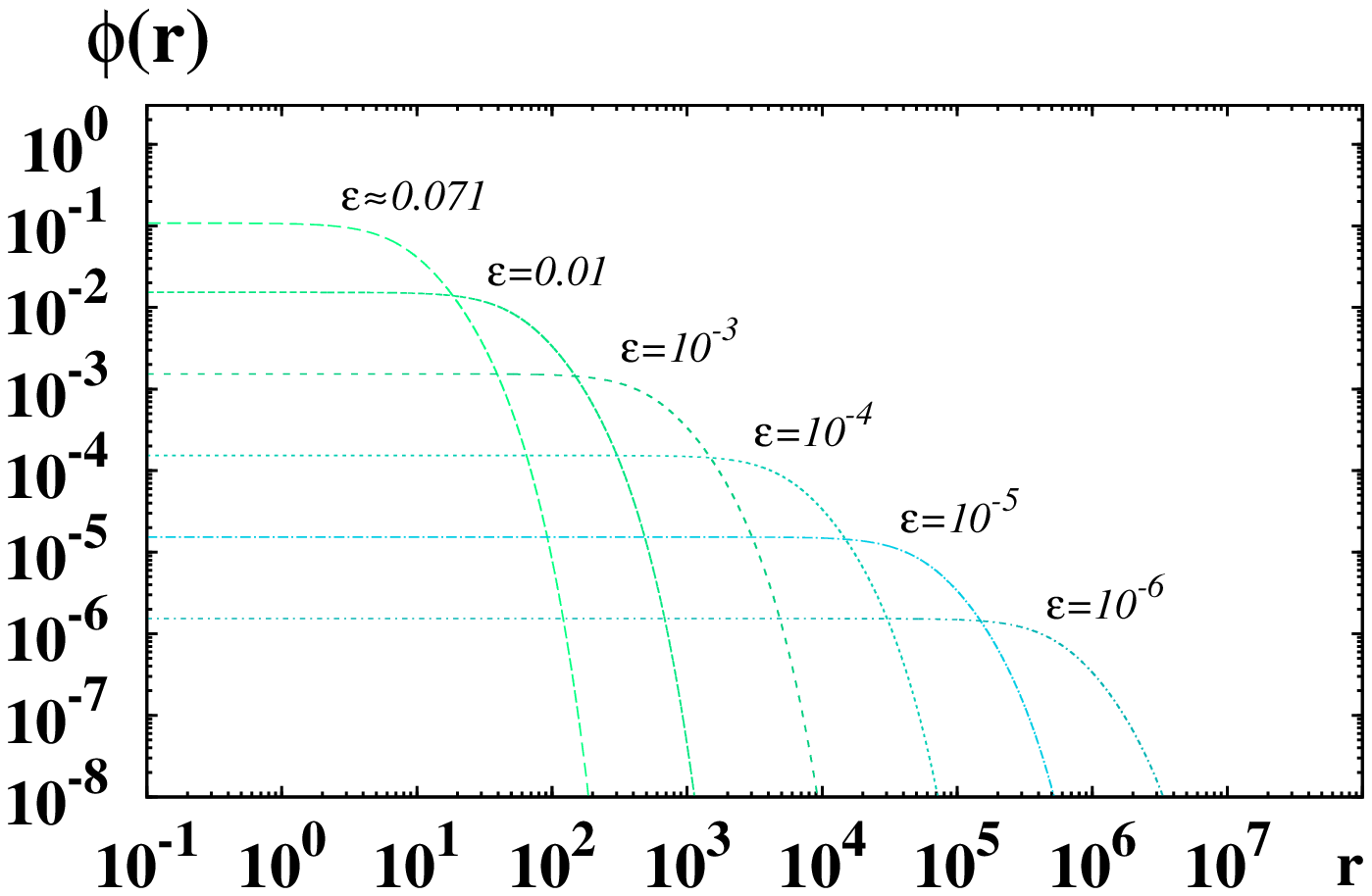}
\includegraphics[width=5.8cm]{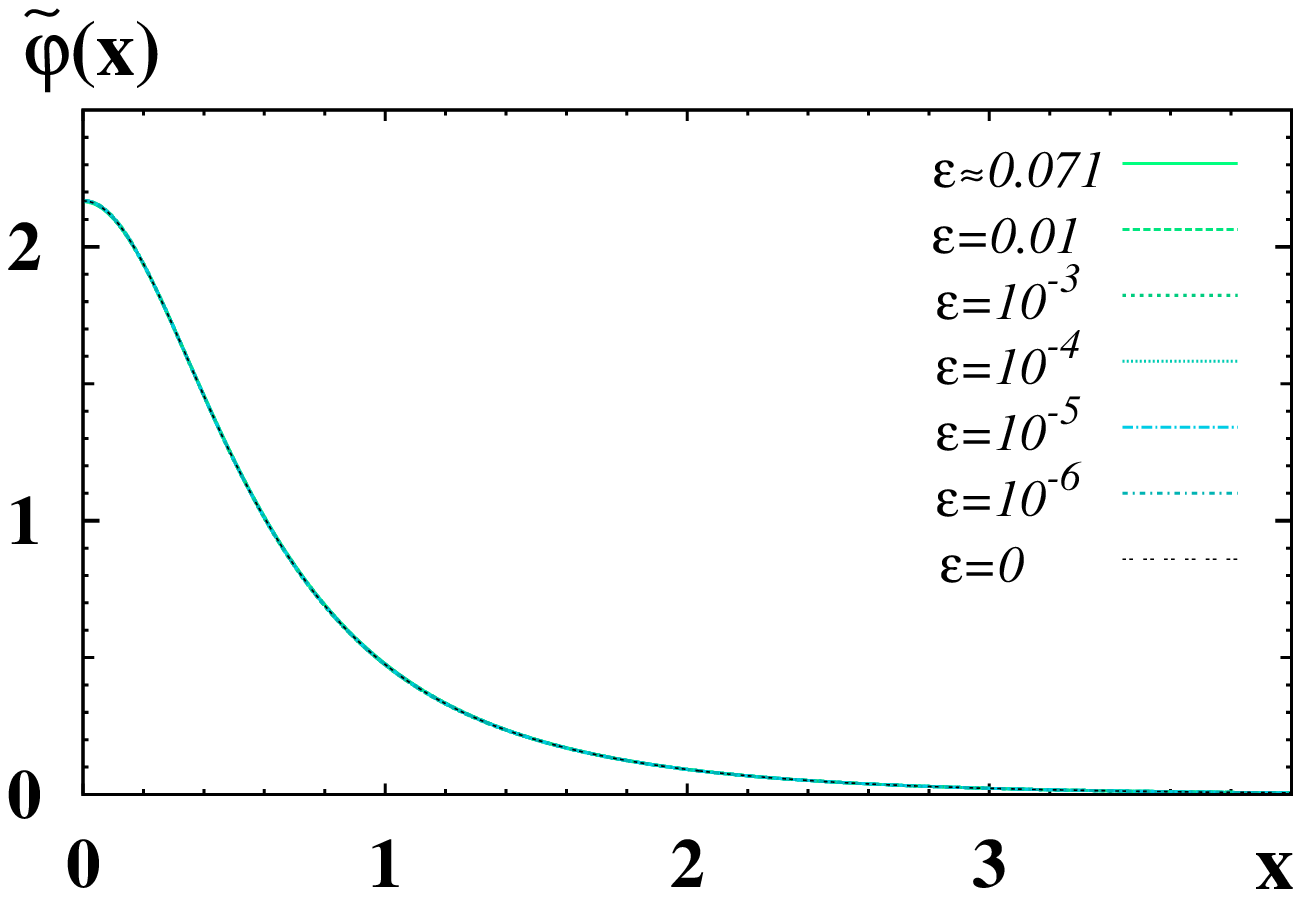}
\caption{\label{FIG-03:phi-r-double-log+rescale}
	(a)
	The solutions $\phi(r)$ as functions of $r$ for different $\eps$
	in a double-logarithmic plot to illustrate how strongly the solutions 
	decrease with $\eps$ and at the same time spread out.
	The result for $\eps\approx0.071$ ($\omega^2=2.195$)
	was the value closest to the $Q$-cloud limit solved in 
	\cite{Mai:2012yc} without the rescaling (\ref{Eq:rescaling}).
	(b)
	Solutions $\rphi(x)$ from (a) rescaled according to
	(\ref{Eq:rescaling}) vs.\ $x$.
	Also the limiting value $\varepsilon=0$ is shown. The 
	different results are indistinguishable within the resolution 
	of this plot.}

\end{figure}

At respectively small and large $x$ the rescaled fields behave as
\ba
    \rphi(x)&=&\rphi_0 + 
    \biggl(\rphi_0-4\rphi_0^3+6\,\alpha\,\teps^2\rphi_0^5\biggr)\,\frac{x^2}{6}
    + {\cal O}(x^4), \nonumber\\
    \rphi(x)&=&\tilde{c}_\infty\;\frac{e^{-x}}{x}\,.
\label{Eq:asymp-resc}
\ea
The 'rescaled problem' (\ref{Eq:eom-resc}) 
can be solved numerically for arbitrarily small $\eps$. 
In Fig.~\ref{FIG-03:phi-r-double-log+rescale}a we show the solutions 
for selected values in the range 
$10^{-6} \le \eps \le 10^{-2}$ for $\phi(r)$ as functions of $r$, 
i.e.\ in the 'usual' units restored according to (\ref{Eq:rescaling}). 
We include the result for $\eps\approx0.071$ corresponding to 
$\omega^2=2.195$ which was the $\omega$-value closest to $\omega_{\rm max}$ 
numerically tractable in \cite{Mai:2012yc}.

Fig.~\ref{FIG-03:phi-r-double-log+rescale}a demonstrates how strongly 
the magnitudes of the solutions decrease with decreasing $\eps$,
and at the same time how much the solutions spread out.

In Fig.~\ref{FIG-03:phi-r-double-log+rescale}b we plot the solutions 
in terms of rescaled fields and coordinates (\ref{Eq:rescaling}).
The rescaled solutions cannot be distinguished from each other for 
$10^{-6} \le \eps \lesssim 10^{-1}$ within the resolution of the plot. 
Deviations from the curves in Fig.~\ref{FIG-03:phi-r-double-log+rescale}b 
start to be noticeable only for $\eps \gtrsim 0.2$, as we will investigate 
in detail in Sec.~\ref{Sec-5:densities}.

We remark that the rescaled problem (\ref{Eq:eom-resc}) allows one 
to extend the study by orders of magnitude into the small-$\eps$ 
region, which is restricted only by numerical accuracy.
In our case the relative accuracy is typically in the range 
$10^{-8}$--$10^{-6}$. We know this from monitoring the numerical 
quality of the solutions by checking, e.g., that 
(\ref{Eq:stability-condition}) holds 
within numerical accuracy, that the different representations 
(\ref{Eq:def-d1-pressure},~\ref{Eq:def-d1-shear}) yield the
same value for $d_1$,  and by performing other quality tests;
see \cite{Mai:2012yc} for more details.
Hence, our numerical method is not sufficiently accurate for  
$\eps < 10^{-6}$. But as is apparent from 
Fig.~\ref{FIG-03:phi-r-double-log+rescale}b, it is not 
necessary to go to such small values of $\eps$.

\begin{figure}[b!]
\includegraphics[width=2.3cm]{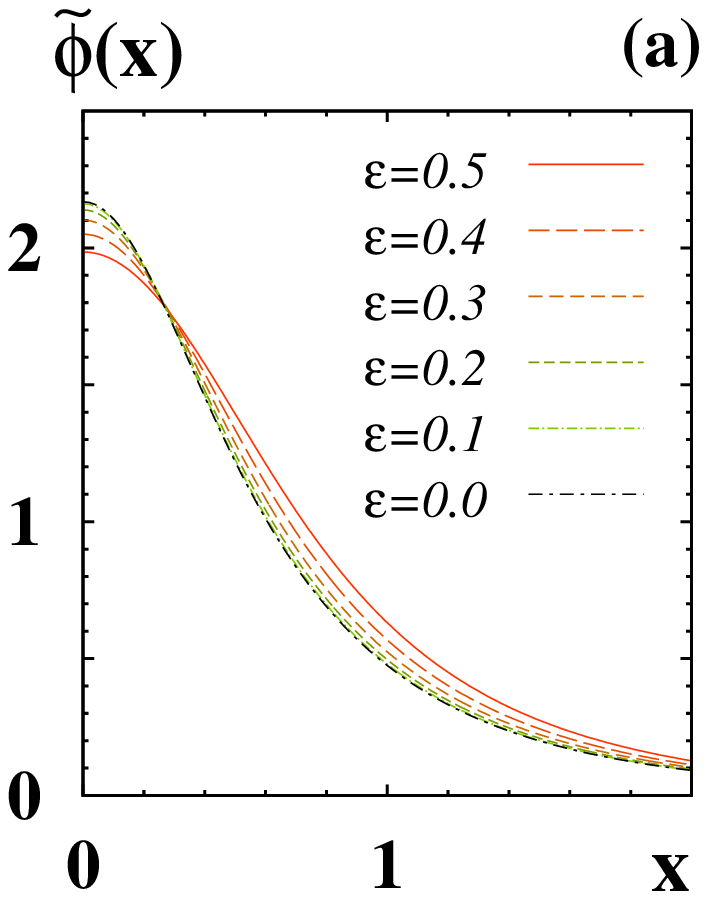} \
\includegraphics[width=2.3cm]{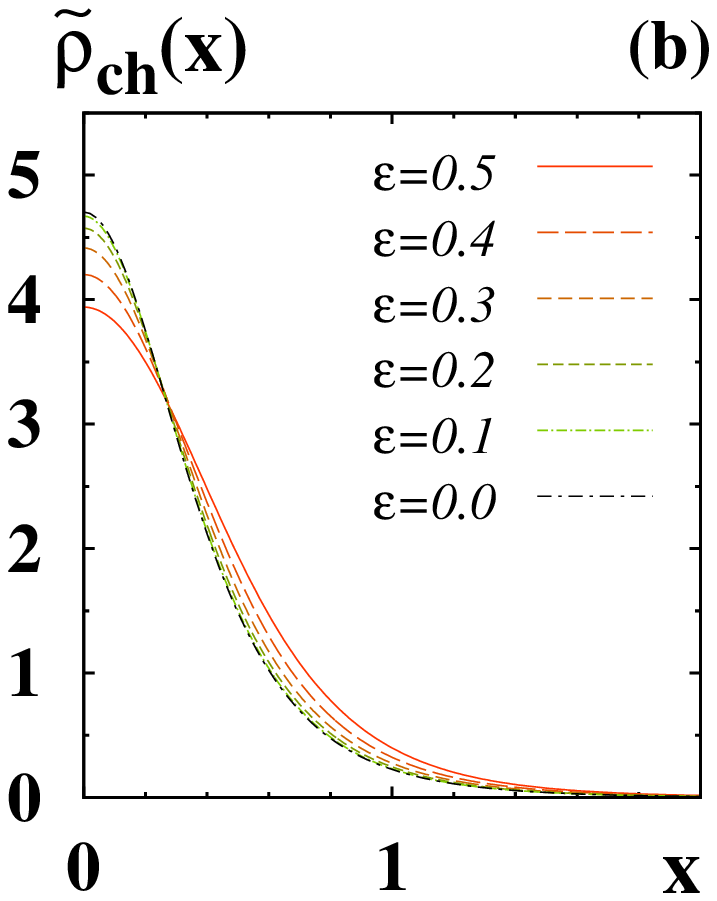} \
\includegraphics[width=2.3cm]{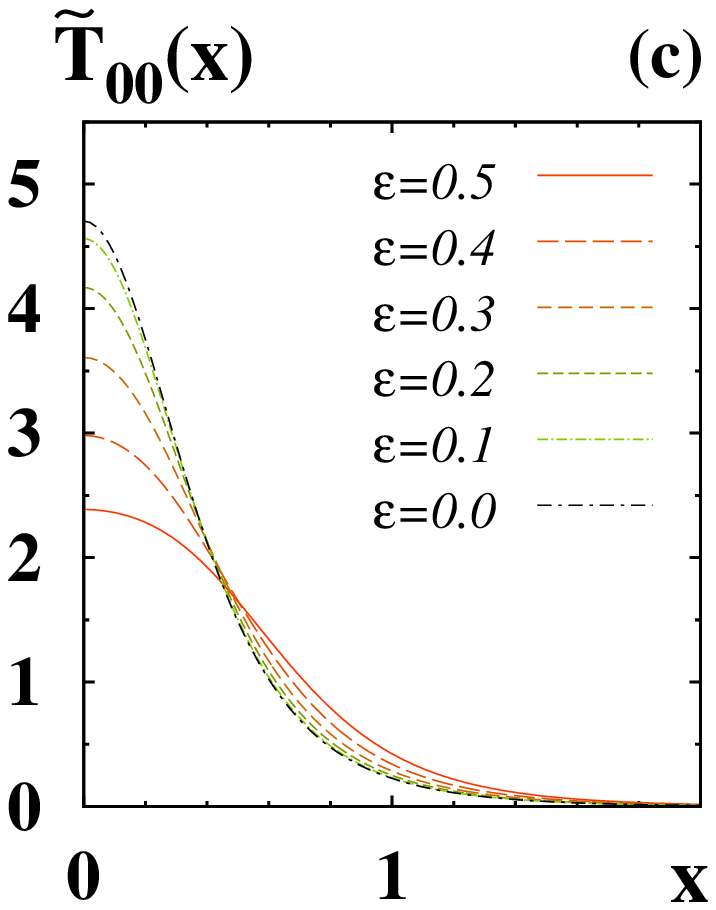} \
\includegraphics[width=2.3cm]{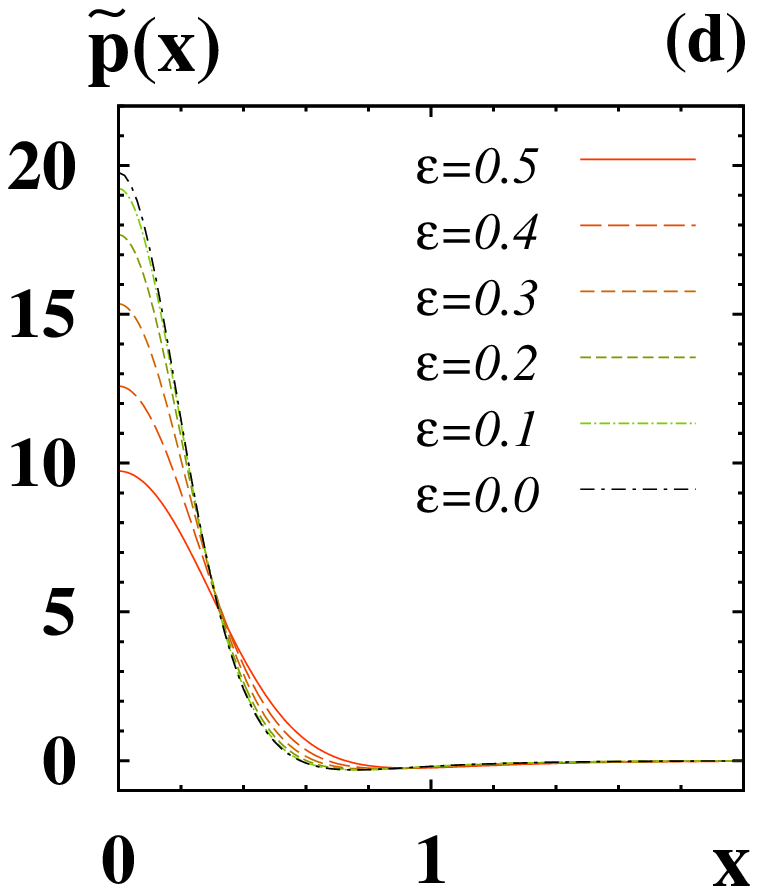}   \
\includegraphics[width=2.3cm]{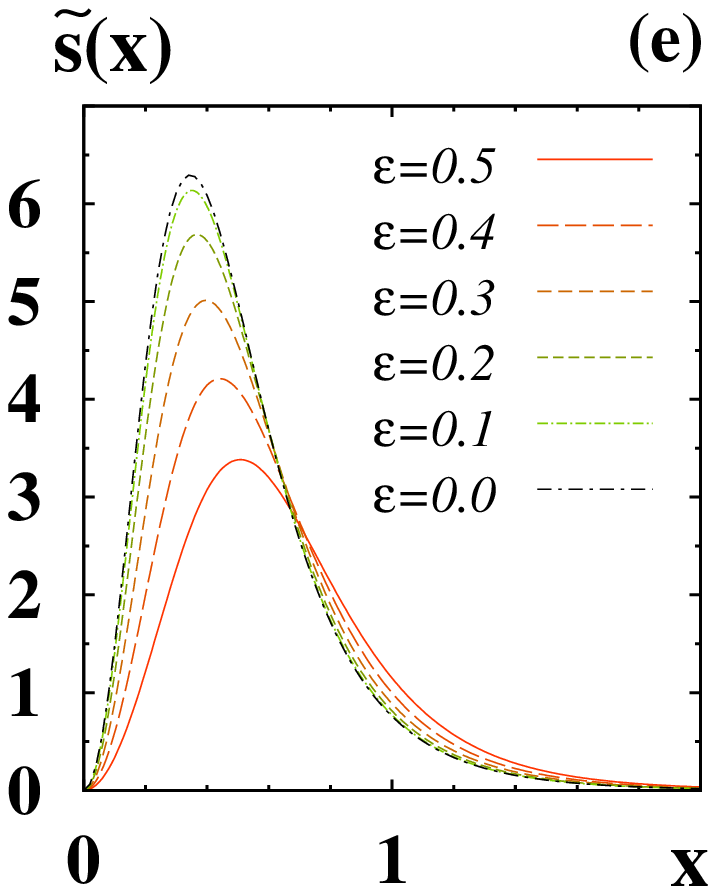} 
\vspace{-2mm}
\caption{\label{Fig-5:rescaled-densities}
	The fields $\rphi(x)$, 
	charge densities $\tilde{\rho}_{\rm ch}(x)$, 
	energy densities $\tilde{T}_{00}(x)$, 
	pressure distributions $\tilde{p}(x)$, 
	shear force distributions $\tilde{s}(x)$
	vs.\ $x$ for selected values of $\eps$.
	The fields and densities are dimensionless 
	after the respective rescaling procedures 
	(\ref{Eq:rescaling},~\ref{Eq:resc-densities}). }
\end{figure}
The physical problem requires 
$\omega^2_{\rm min} < \omega^2 < \omega^2_{\rm max}$ which implies
$\teps>0$, see Eqs.~(\ref{Eq:condition-for-existence},
\ref{Eq:teps}). But in the rescaled problem 
(\ref{Eq:eom-resc}) nothing prevents us from setting $\teps = 0$.
This must be understood as a careful limiting
procedure, otherwise the rescaling (\ref{Eq:rescaling}) would 
become singular. 

Nevertheless, in the limit $\teps = 0$ the problem
(\ref{Eq:eom-resc}) has a regular solution $\rphi(x)$ which we 
have included in  Fig.~\ref{FIG-03:phi-r-double-log+rescale}b,
and which also cannot be distinguished from the curves
plotted in Fig.~\ref{FIG-03:phi-r-double-log+rescale}b which 
refer to finite $\eps$ in the range $10^{-6}\le \eps\lesssim 0.1$.
This shows how smoothly the limit is approached, and 
how numerically close to it the results are for finite 
$\eps\lesssim 0.1$.

\section{Densities in the limit \boldmath $\eps\to 0$}
\label{Sec-5:densities}

To investigate the behavior and properties of the solutions 
in the limit $\eps\to0$, one has to go to more sizable values of
$\eps > 0.1$. In the following, in order to visualize how the 
rescaled fields $\rphi(x)$ approach the limiting case, we will 
present results for $\eps=0, \, 0.1, \, 0.2, \, 0.3, \, 0.4, \, 0.5$.

In Fig.~\ref{Fig-5:rescaled-densities}a we show the rescaled 
fields $\rphi(x)$ as functions of $x$ for different $\eps$.
Despite this sizable $\eps$-range, the variation of 
$\rphi(x)$ with decreasing $\teps$ is modest. 
As the limit is approached, the solutions tend to grow in the 
center region and decrease at larger $x$. As a result, the 
limiting field is more strongly localized around $x=0$ 
than the solutions for finite $\eps$.

To discuss further properties, we define the rescaled 
densities as follows (see App.~\ref{App:systematic-notation} 
for alternative notation)
\begin{align}
\rho_{\rm ch}(r)	&= \eps^2\;\frac{\omega}{B}\;\tilde{\rho}_{\rm ch}(x), &
   p(r)		&= \frac{\eps^4}{B}\;\tilde{p}(x), \nonumber\\
   T_{00}(r)	&= \eps^2\;\frac{2A}{B}\;\tilde{T}_{00}(x), &
   s(r) 	&= \frac{\eps^4}{B}\;\tilde{s}(x), 
  \label{Eq:resc-densities}
\end{align}
with
\begin{align}
	\tilde{\rho}_{\rm ch}(x)	&= \rphi(x)^2\,,\hspace{-15mm}&& \nonumber\\
	\tilde{T}_{00}(x)	&= \rphi(x)^2   
	+ \teps^2\biggl[\hspace{-15mm}&&\frac12\,\rphi^\prime(x)^2 
	- \frac12\,\rphi(x)^2-\rphi(x)^4
	+ \alpha\,\teps^2\rphi(x)^6\biggr],  \nonumber\\ 
 	\tilde{p}(x) 	& = \hspace{-15mm}&-&{ }\frac16\,\rphi^\prime(x)^2
	-\frac12\,\rphi(x)^2+\rphi(x)^4 
	-\alpha\,\teps^2\rphi(x)^6, \nonumber\\ 
 	\tilde{s}(x) 	&= \hspace{-15mm}&& \hspace{3.5mm}\rphi^\prime(x)^2 \, .
 \label{Eq:resc-densities-def}	
\end{align}
The rescaling is such that physical dimensions, including leading 
powers of $\eps$, are stripped off in (\ref{Eq:resc-densities}),
such that the tilde-densities  in Eq.~(\ref{Eq:resc-densities-def})
are expressed solely in terms of dimensionless parameters and fields 
$\alpha$, $\teps$, $\rphi(x)$.

Interestingly, in the limiting case, and only in this case,
the rescaled charge and energy densities coincide. 
The differential equation (\ref{Eq:diff-eq-s-p}) connecting 
$p(r)$ and $s(r)$ holds analogously for $\tilde{s}(x)$ and
$\tilde{p}(x)$, and is satisfied for $\teps>0$ as well as 
in the limit $\teps=0$. The same holds true for the
von Laue condition (\ref{Eq:stability-condition}).
We will follow up on the meaning of these observations
in Sec.~\ref{Sec-7:interpretation}.

In Fig.~\ref{Fig-5:rescaled-densities}b and
Fig.~\ref{Fig-5:rescaled-densities}c we show 
the rescaled charge and energy densities. Both densities 
grow with decreasing $\eps$ in the center region, decrease 
in the outer region, and coincide for $\eps\to 0$.
But $\tilde{T}_{00}(x)$ shows a stronger dependence on $\eps>0$
than $\tilde{\rho}_{\rm ch}(x)$ because, in contrast to  charge 
density, the energy density encodes the full information on the
dynamics of the theory. 

The rescaled pressure function $\tilde{p}(x)$ is shown in 
Fig.~\ref{Fig-5:rescaled-densities}d. As the limit is approached, 
$\tilde{p}(x)$ grows in the center and decreases in the outer region. 
For all solutions, including the limiting case, the rescaled pressure 
exhibits precisely one zero,  as all ground state 
$Q$-ball solutions do \cite{Mai:2012yc}.

In Fig.~\ref{Fig-5:rescaled-densities}e we show the rescaled shear 
force distribution. In the opposite limit $\omega\to\omega_{\rm min}$ 
where $Q$-balls share some features of liquid drops \cite{Mai:2012yc}, 
the shear force distributions approach the shape of a $\delta$-function 
concentrated around the ``sharp edge'' of the $Q$-balls. In the $Q$-cloud 
limit $\omega\to\omega_{\rm max}$ the positions of the peaks of $\tilde{s}(x)$ 
still indicate the size of the solutions. But the considerable widths 
of the $\tilde{s}(x)$ imply that the solutions have no ``sharp edge'' 
(not even in the rescaled coordinates $x$) but are diffuse.

\begin{figure}[b!]
\vspace{-2mm}

\begin{center}
\includegraphics[height=3.3cm]{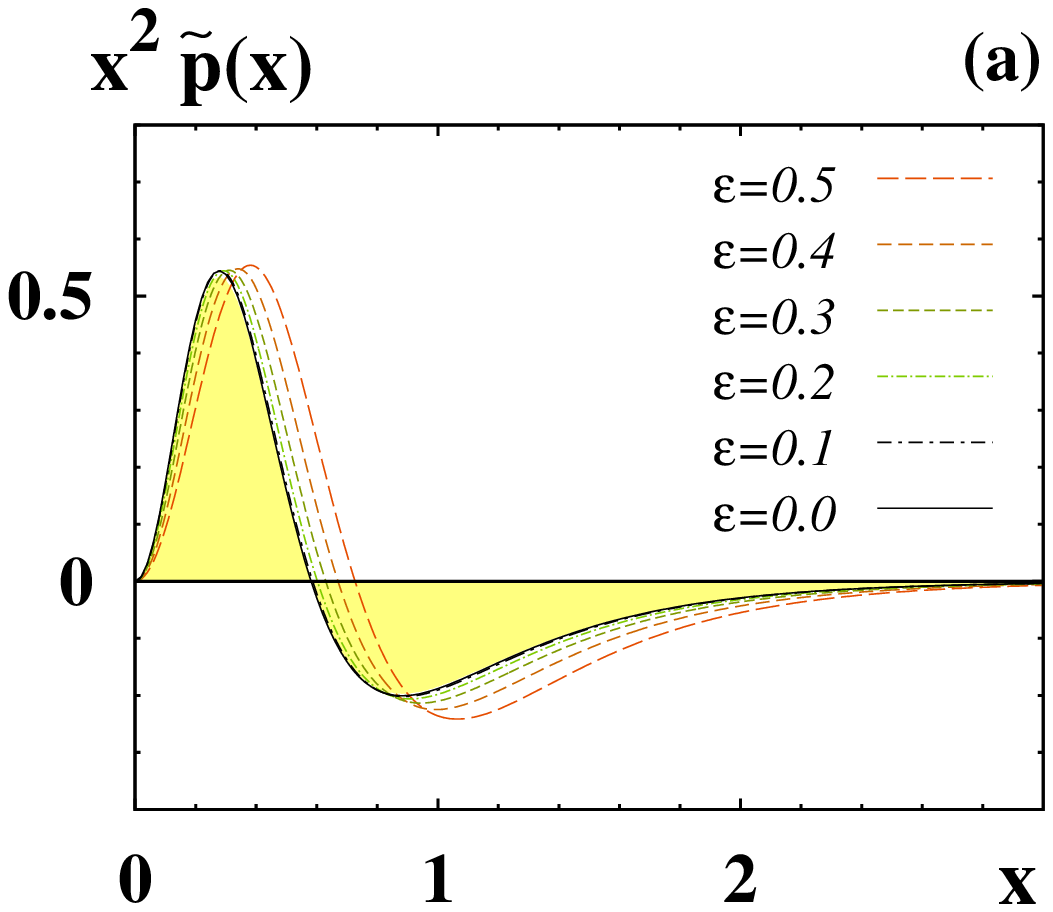} \
\includegraphics[height=3.3cm]{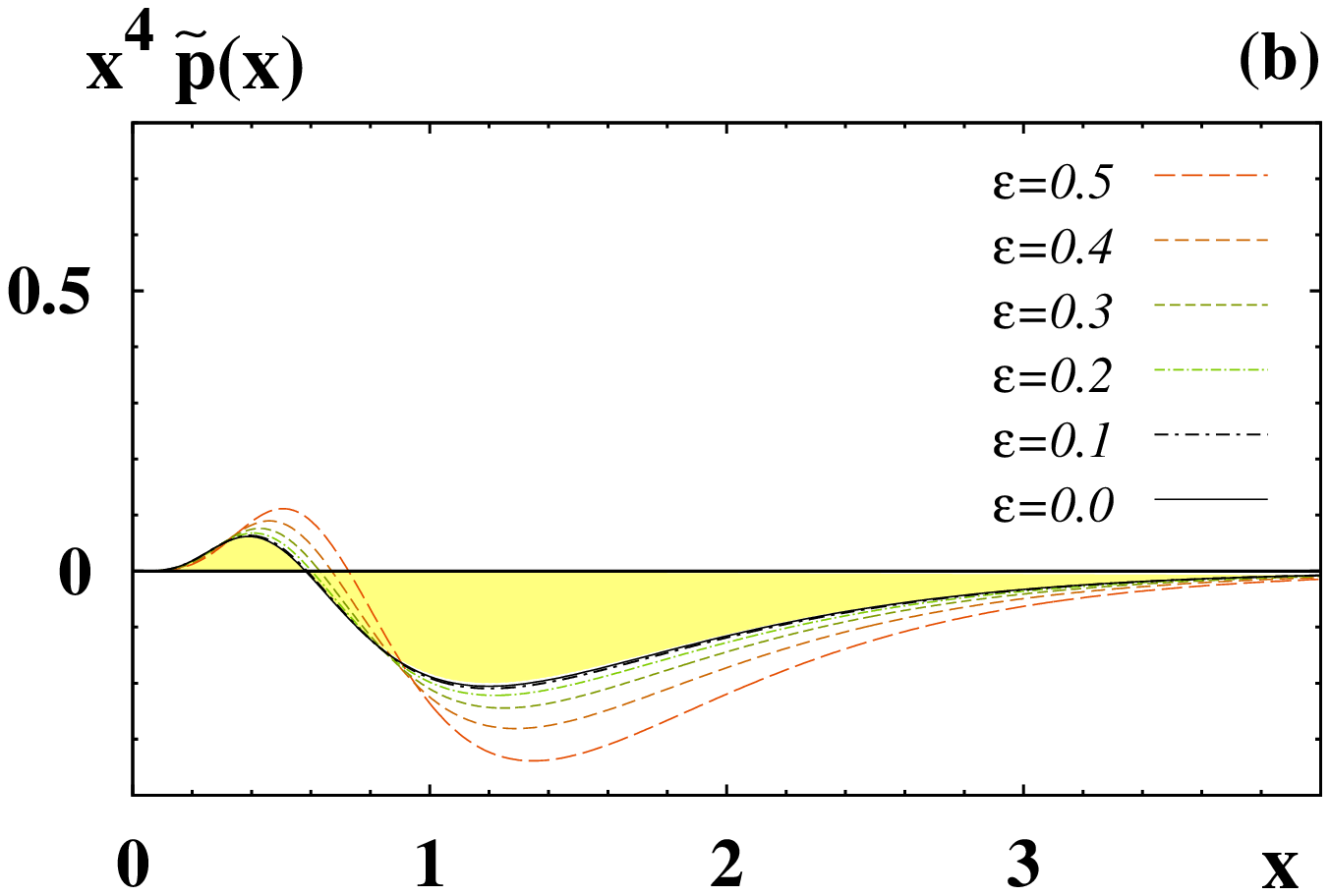} 
\end{center}

\vspace{-2mm}
\caption{\label{Fig-5:rescaled-p-x2-x4}
	$\tilde{p}(x)$ weighted with $x^2$ and $x^4$ vs.\ $x$ 
	for selected $\eps$.
	Integrating $x^2\tilde{p}(x)$ yields zero, 
	Eq.~(\ref{Eq:stability-condition}).
	Integrating $x^4\tilde{p}(x)$ yields a negative result,
	which explains that $d_1<0$, see Eq.~(\ref{Eq:def-d1-pressure}).
	For $\eps=0$ the areas under the curves are shaded.}
\end{figure}

In Fig.~\ref{Fig-5:rescaled-p-x2-x4} we plot the results for 
$x^2\tilde{p}(x)$, $x^4\tilde{p}(x)$ which shows more clearly the
zeros of the pressure distribution.
The results for $x^2\tilde{p}(x)$ in 
Fig.~\ref{Fig-5:rescaled-p-x2-x4}a illustrate the balance of the 
internal forces. Positive (negative) pressure in center (outer)
region corresponds to repulsion (attraction).
Repulsive and attractive forces balance each other exactly 
according to (\ref{Eq:stability-condition}).
The areas under the curves in Fig.~\ref{Fig-5:rescaled-p-x2-x4}a
are equal within numerical accuracy for all $\eps$.

The results for $x^4\tilde{p}(x)$ in Fig.~\ref{Fig-5:rescaled-p-x2-x4}b
visualize the integrand of $d_1$ in (\ref{Eq:def-d1-pressure}).
Since $x^2\tilde{p}(x)$ integrates to zero, the integral
of $x^4\tilde{p}(x)$ is negative as the additional weight $x^2$
diminishes the contribution from the positive inner
region and enhances that of the negative outer region.
The same pattern was observed in other models, and illustrates 
the relation of $d_1$ to internal forces in the system.

This shows that the $D$-term of $Q$-clouds is negative. 
The sign of $d_1$ can be concluded independently and much more 
directly by exploring Eq.~(\ref{Eq:def-d1-shear}) which relates 
$d_1$ to the positive definite shear forces (\ref{Eq:shear}).
In consistent and correctly solved field-theories one can use 
both (\ref{Eq:def-d1-pressure})  and (\ref{Eq:def-d1-shear})  to 
show that $d_1<0$. In fact, for regular $Q$-ball solutions the 
relation (\ref{Eq:def-d1-shear}) shows immediately --- without 
numerical calculations --- that $d_1<0$ \cite{Mai:2012yc,Mai:2012cx}.
However, concluding the sign of $d_1$ from (\ref{Eq:def-d1-pressure}) 
allows one to verify the von Laue condition (\ref{Eq:stability-condition}) 
which provides a cross check whether the equations of motion 
(here the minimization of the soliton energy) 
have been correctly solved.
This is of particular importance in our case,
since the limiting solution is strictly speaking singular
and many properties, including the $D$-term, diverge such 
that one cannot take the equivalence of the representations
(\ref{Eq:def-d1-pressure})  and (\ref{Eq:def-d1-shear}) 
for $d_1$ for granted.

\section{Global properties for \boldmath $\eps\to 0$}
\label{Sec-6:global-properties}

The global properties of the solutions, which follow from integrating the 
densities, can be expressed as
(the notation is such that the left-hand sides 
in (\ref{Eq:scaling-global-properties}) have well-defined 
finite limits for $\eps\to 0$)
\begin{subequations}\label{Eq:scaling-global-properties}
\begin{alignat}
   1\eps\,Q &=\frac{\omega}{B}\,\int\!\di^3x\;\tilde{\rho}_{\rm ch}(x),
    \label{Eq:scaling-Q}\\
   \eps\,M &=\frac{2A}{B}    \,\int\!\di^3x\;\tilde{T}_{00}(x),     
   \label{Eq:scaling-M}\\
   \eps^2d_1&= \frac{5(\eps M) }{4B}\,\int\!\di^3x\;x^2\tilde{p}(x)
    	    =  -\,\frac{(\eps M) }{3B}\,\int\!\di^3x\;x^2\tilde{s}(x), 
   \label{Eq:scaling-d1}\\
   \eps^2 \la r^2_s\ra 
	&=\frac{\int_0^\infty\di x\;x^2\tilde{s}(x)}
                {\int_0^\infty\di x\;   \tilde{s}(x)}, \label{Eq:scaling-rs}\\
   \eps^2 \la r^2_Q\ra 
	&=\frac{\int\!\di^3x\;x^2\tilde{\rho}_{\rm ch}(x)}
                {\int\!\di^3x\;\tilde{\rho}_{\rm ch}(x)}, \label{Eq:scaling-rQ}\\
   \eps^2 \la r^2_E\ra 
	&=\frac{\int\!\di^3x\;x^2\tilde{T}_{00}(x)}
                {\int\!\di^3x\;\tilde{T}_{00}(x)} \label{Eq:scaling-rE}\,,\\
   \eps^{-3}\gamma &= \frac{1}{B}\int_0^\infty\di x \,\tilde{s}(x)\,.
   \label{Eq:scaling-gamma}
\end{alignat}
\end{subequations}

From (\ref{Eq:scaling-global-properties}) we expect that with
decreasing $\eps$ the charge, mass and mean radii diverge 
as $1/\eps$ and $d_1$ grows as $1/\eps^2$, while the surface tension 
$\gamma$ vanishes as $\eps^3$. 
Another relevant property is surface energy which scales as
\begin{align}\begin{subequations}
   	\eps^{-1}\,E_{\rm surf} 
	=\frac{1}{B}\,\int\!\di^3 x\;\tilde{s}(x).
   	\tag{\ref{Eq:scaling-global-properties}h}
	\label{Eq:scaling-Esurf}
\end{subequations}\end{align}
where $E_{\rm surf}$ can be defined in equivalent ways as \cite{Mai:2012yc}
(here $D=3$ denotes the number of space dimensions),
\setcounter{equation}{32}
\be\label{Eq:define-Esurf}
	E_{\rm surf} 
	 = 4\pi\,\gamma\,\la r_s^2\ra 
	\equiv D\,(M-\omega Q) \,.
\ee
Thus $E_{\rm surf}\propto\eps$ for $\eps\to0$. We will 
need this result later. 

We recall that for $\omega\to\omega_{\rm min}$ $Q$-balls develop a 
``sharp edge'' which makes $\gamma$ and $E_{\rm surf}$ well-defined 
notions \cite{Coleman:1985ki}. It is interesting 
that in the opposite limit $\omega\to\omega_{\rm max}$ 
surface tension and surface energy become irrelevant.

\begin{figure}[t!]
\centering
\includegraphics[width=3.1cm]{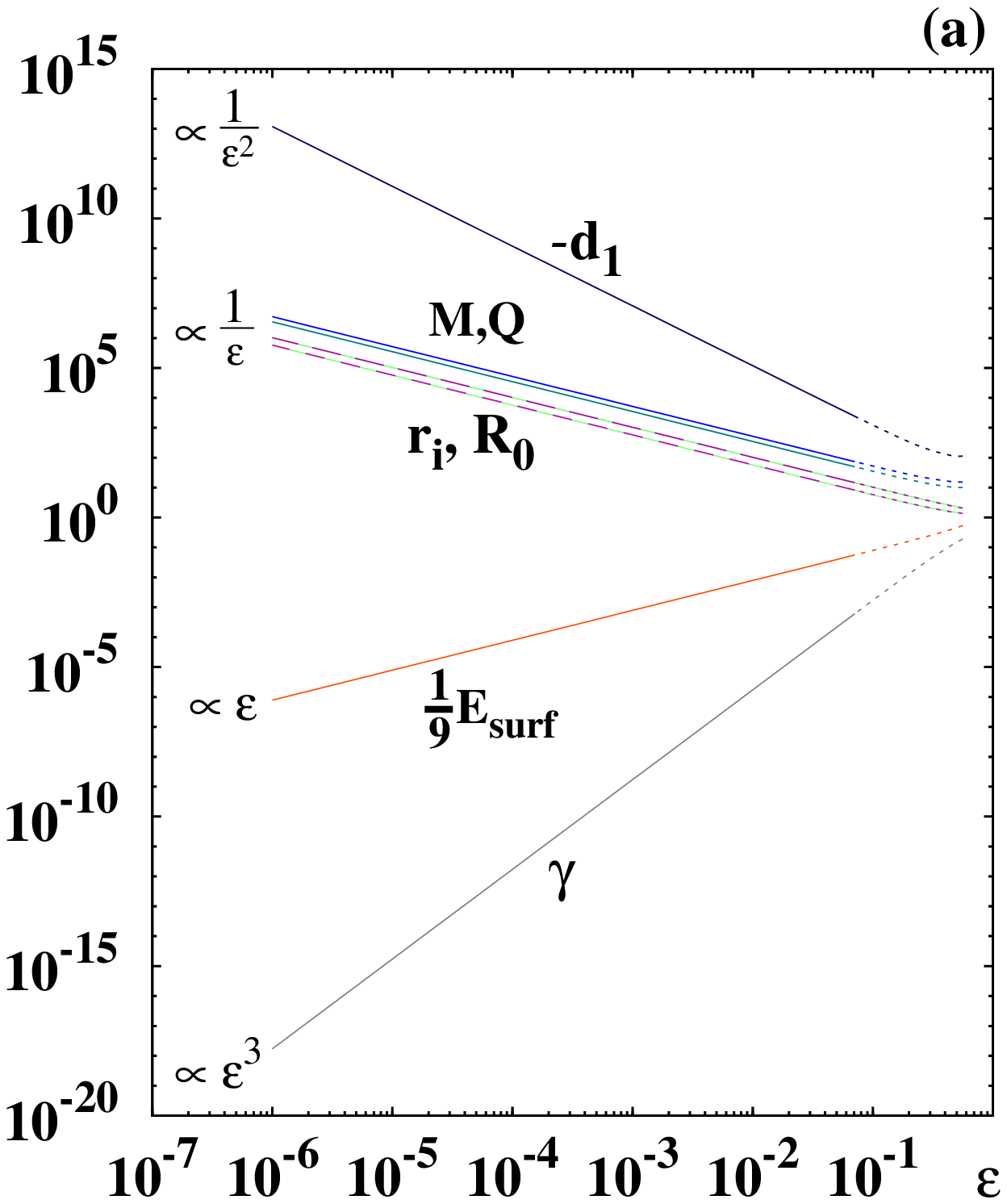} \ 
\includegraphics[width=3.1cm]{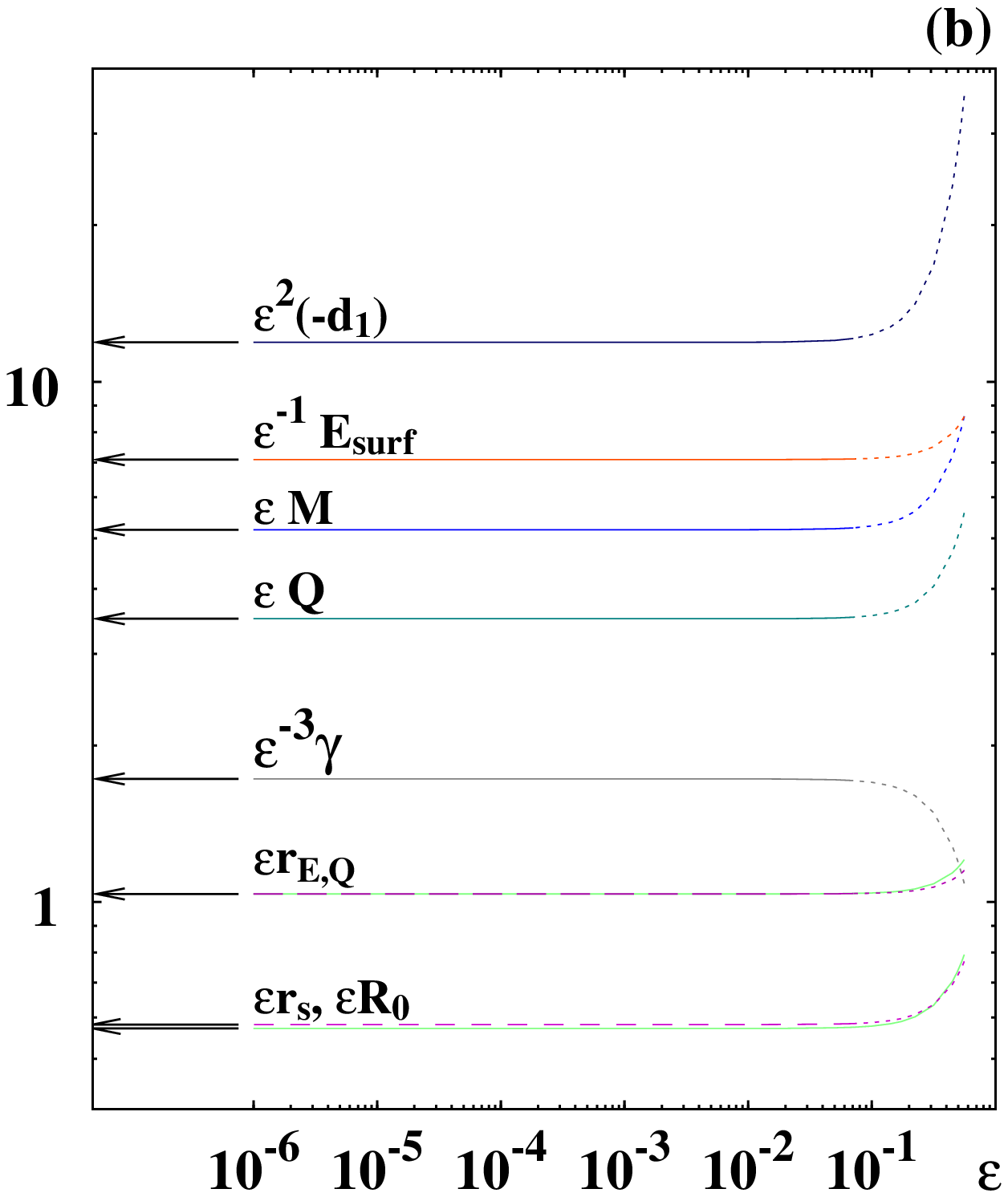} \ \ \ 
\includegraphics[width=3.1cm]{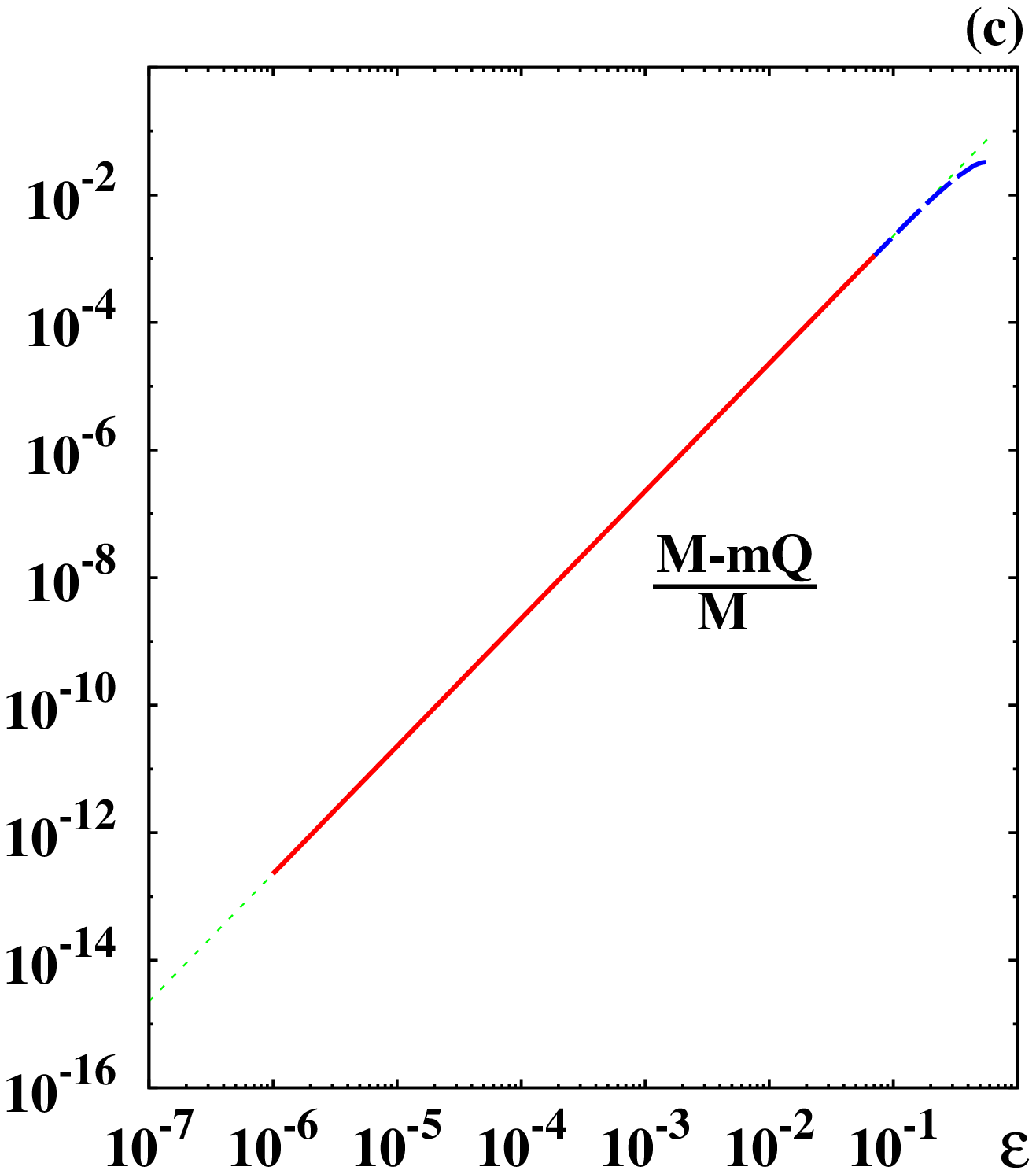} 
\caption{\label{Fig-6:Q-cloud-limit}
  (a) $Q$-ball properties as functions of $\eps$.
  From top to bottom: 
  $d_1$, $M$, $Q$, $r_i=\la r^2_i\ra^{1/2}$ ($i=E,\,Q,\,s$), $R_0$,
  $E_{\rm surf}$, $\gamma$. 
  Dotted curves corresponding to $1.8\le\omega^2\le 2.195$
  are from \cite{Mai:2012yc},
  solid or long-dashed curves were obtained in this work.
  We plot $\frac19\,E_{\rm surf}$ to shift the curve and avoid 
  intersections with other curves.
  (b) The same as (a) but for scaled properties.
  (c) $(M-mQ)$ normalized with respect to $M$ vs.~$\eps$.
  Solid (dashed) lines: results obtained here (in Ref.~\cite{Mai:2012yc})
  as described in caption of Fig.~\ref{Fig-6:Q-cloud-limit}.
  Dotted line: the analytically calculated result from 
  Eq.~(\ref{Eq:limit-M-vs-mQ}) with higher order terms 
  ${\cal O}(\teps^4)$ neglected.}
\vspace{-2mm}
\end{figure}

In Fig.~\ref{Fig-6:Q-cloud-limit}a we show numerical results 
for the properties $d_1$, $M$, $Q$, $r_i=\la r^2_i\ra^{1/2}$ 
($i=E,\,Q,\,s$), $R_0$, $E_{\rm surf}$ and $\gamma$. Here $R_0$ 
denotes the position of the zero of the pressure, 
i.e.\ $p(R_0)=0$. We include the results from \cite{Mai:2012yc} 
(dotted lines) for $1.8\le\omega^2\le 2.195$ corresponding to 
$0.071 \lesssim \eps \lesssim 0.63$. The results obtained here
extend the work of \cite{Mai:2012yc} down to $\eps=10^{-6}$
(solid lines).  
We stress that we obtain within numerical accuracy the same result 
for $d_1$ using Eq.~(\ref{Eq:shear}) or (\ref{Eq:def-d1-pressure}).

On a plot with the resolution of Fig.~\ref{Fig-6:Q-cloud-limit}a
the curves for $\la r^2_E\ra^{1/2}$ and $\la r^2_Q\ra^{1/2}$ are
practically on top of each other, which is due to the fact
that both radii coincide for $\eps\to 0$. 
In Fig.~\ref{Fig-6:Q-cloud-limit}a also the curves for 
$\la r^2_s\ra^{1/2}$ and $R_0$ are practically indistinguishable,\footnote{
	We use the occasion to correct 2 details in Fig.~11 of 
	\cite{Mai:2012yc}. In that figure the results for $\la r_s^2\ra^{1/2}$ 
	were reported to be on top of the curves for $\la r_E^2\ra^{1/2}$ 
	and $\la r_Q^2\ra^{1/2}$, while they were actually on top of $R_0$ 
	(as in this work). In addition, the shift of the curve for $E_{\rm surf}$ 
	by the factor $\frac19$ was not mentioned in the caption of 
	Fig.~11 of \cite{Mai:2012yc}. These corrections have no 
	consequence on the results and conclusions of \cite{Mai:2012yc}.} 
but this effect is accidental due to the logarithmic scale in the plot 
because $R_0$ and $\la r^2_s\ra^{1/2}$ are numerically close but not equal.

In Fig.~\ref{Fig-6:Q-cloud-limit}b we show the behavior of the 
adequately scaled quantities as defined on the left-hand sides 
in (\ref{Eq:scaling-global-properties}). The arrows indicate
the results in the limiting case $\eps=0$. The figure illustrates
that the scaling regime practically sets in for $\eps\lesssim 0.1$,
i.e.\ it could just be observed in \cite{Mai:2012yc}.

Next we discuss how $M$ approaches $mQ$ for $\eps\to0$. 
For $\omega>\omega_c$ (i.e.\ for $\eps \lesssim 0.55$
for the parameters used in this work) one has $M>mQ$,
see Sec.~\ref{Sec-2:EMT-and-Qballs}, such that $M$ approaches 
$mQ$ from above \cite{Alford:1987vs}. 
We can make a more quantitative statement by evaluating the leading 
term of $(M-mQ)/M$. Remarkably, the result can be computed analytically,
see App.~\ref{App:limit-M-vs-mQ}, 
\be\label{Eq:limit-M-vs-mQ}
	\frac{M-mQ}{M} = \frac{1}{2}\;\teps^2 + {\cal O}(\teps^4)\,.
\ee
In Fig.~\ref{Fig-6:Q-cloud-limit}c we show $(M-mQ)/M$ down to 
$\eps \ge 10^{-6}$ using the numerical results for $M$ and $Q$ 
displayed in Fig.~\ref{Fig-6:Q-cloud-limit}.
We include the analytical result (\ref{Eq:limit-M-vs-mQ}) 
neglecting higher order terms ${\cal O}(\teps^4)$. 
In the regime $\eps < 10^{-2}$ the numerical results practically coincide 
with the leading order asymptotics of Eq.~(\ref{Eq:limit-M-vs-mQ}).

\section{Interpretation}
\label{Sec-7:interpretation}

Having established the behavior of the properties of the solutions in 
the limit $\eps\to0$, we now turn to the interpretation of the results. 
One way to understand the limiting solution consists of
interpreting it as a dissociation of the unstable solutions into an 
infinitely dilute system of uniformly distributed free $Q$-quanta, 
a ``$Q$-cloud'' \cite{Alford:1987vs}.

That the system becomes dilute can be seen from the densities 
(\ref{Eq:resc-densities},~\ref{Eq:resc-densities-def}).
At every point in space the charge and energy densities vanish as 
$\eps^2$, which shows that the system is ``infinitely dilute.''
At the same time the internal forces characterized by
$p(r)$ and $s(r)$ vanish even faster as $\eps^4$.
This implies that in the $Q$-cloud limit the interactions 
between the quanta decrease, and they become free.
That in the limit the $Q$-quanta can be considered free may
be inferred alternatively from the observation that in the
limit the rescaled charge and energy densities become
equal, see Eq.~(\ref{Eq:resc-densities-def}), implying that $M=m\,Q$.
The conserved charge $Q$ ``counts'' the number of quanta. 
Thus, the total mass of the system is given  by the 
number of elementary quanta multiplied by their mass $m$. 
This in turn implies that the binding energy vanishes,
i.e.\  the quanta are free in the limit.
(As the absolute stability condition $M<m\,Q$ is not
satisfied for $\omega_{\rm abs} < \omega < \omega_{\rm max}$,
the limit of a free gas of quanta is approached from above 
through the regimes of meta- and unstable solutions.)
The last statement to clarify is the ``uniform'' distribution
of the free quanta. This is explained by considering 
Fig.~\ref{FIG-03:phi-r-double-log+rescale}a. 
The solutions, and hence their charge or energy densities, are 
practically constant functions of $r$ up to distances of 
order $1/\eps$. For a sufficiently small $\eps$ one could 
envision the ``visible universe'' filled with a uniformly 
distributed dilute gas of $Q$-quanta.

The results obtained in this work suggest also an alternative
interpretation. For that let us first investigate how the relation
of $M$ and $m\,Q$ arises.
A powerful tool to provide insight in this respect is the virial 
theorem \cite{Kusenko:1997ad}.
This theorem is derived exploring Eqs.~(\ref{Eq:charge},~\ref{Eq:T00}) 
which allow us to express the mass $M$, for a fixed charge $Q$, as
\be\label{Eq:virial-theorem-I}
      M=\frac12\,E_{\rm ch}
       +\frac12\,E_{\rm surf}
       + E_{\rm pot}
\ee
where (keeping the number of dimensions $D=3$ general)
\ba\label{Eq:virial-theorem-Ib}
     E_{\rm surf}=\int\di^Dr\,\phi^\prime(r)^2\;,&&
     E_{\rm pot }=\int\di^Dr\,V(\phi)\;,\nonumber\\
                I=\int\di^D r\,\phi(r)^2\;,&&
     E_{\rm ch}=\frac{Q^2}{I} \;. \label{Eq:def-Esurf-Ech-Epot}
\ea
Now we define $M(\lambda)$ by evaluating (\ref{Eq:virial-theorem-I}) 
for dilatational variations of the solutions 
$\phi(r)\to\phi(\lambda r)$ with $\lambda > 0$, and
substituting $\vec{r}\to\lambda \vec{r}$ in the integrals in
(\ref{Eq:virial-theorem-Ib}). We obtain
\ba\label{Eq:virial-theorem-II}
      M(\lambda)=\frac12\,E_{\rm ch}\;\lambda^D
      +\frac12\,E_{\rm surf} \;\lambda^{2-D}
      +E_{\rm pot} \;\lambda^{-D} \;. \\ \nonumber
\ea
For $\lambda=1$ one has $M^\prime(\lambda)=0$ and $M^{\prime\prime}(\lambda)>0$
since, when setting $\lambda$ to unity, we restore the solutions 
$\phi(r)$ which minimize the energy functional. 
The statement $M^\prime(\lambda)=0$ at $\lambda=1$ is often referred 
to as virial theorem. From this relation one can derive the von 
Laue condition (\ref{Eq:stability-condition}) \cite{Mai:2012yc}. 
Another usage of the virial theorem is to eliminate 
$E_{\rm pot}$ in $M$ which yields the second expression for 
$E_{\rm surf}$ in Eq.~(\ref{Eq:define-Esurf}).

It is instructive to express $M(\lambda)$ in terms of rescaled fields 
(\ref{Eq:rescaling}). (Notice that for $\eps\neq0$ the substitutions 
$\vec{r}\to \lambda  \vec{r}$, $\vec{r}\to\eps\vec{x}$ commute with 
$\vec{r}\to \eps\vec{x}$, $\vec{x}\to\lambda\,\vec{x}$.)
Inserting (\ref{Eq:resc-densities},~\ref{Eq:resc-densities-def})
in (\ref{Eq:virial-theorem-II}) we obtain
\ba\label{Eq:M-lambda}
	\eps M(\lambda)
	&=&\frac{2A}{B}\int\di^Dx
	\Biggl\{
	  \frac{\lambda^D+\lambda^{-D}}{2}\,\rphi(x)^2 \\
	&+& \teps^2\,\Biggl[
		 \frac{\lambda^{2-D}}{2}\,\rphi^\prime(x)^2\,
		-\,\frac{\lambda^D}{2}\,\rphi(x)^2\,
		-\lambda^{-D}\rphi(x)^4\Biggr]
		+\lambda^{-D}\alpha\,\teps^4\,\rphi(x)^6
	\Biggr\} \,. \nonumber
\ea

Setting $\lambda=1$ in (\ref{Eq:M-lambda}) we recover the expression
which follows directly from integrating $T_{00}(r)$ in 
(\ref{Eq:resc-densities}), namely 
\be
	\eps M(\lambda)|_{\lambda=1} \equiv 
	\eps M=\frac{2A}{B}\,\int\di^Dx\;\tilde{T}_{00}(x)\,.
\ee
From $\eps M^\prime(\lambda)$ at $\lambda=1$
%
%
we obtain the von Laue condition (\ref{Eq:stability-condition}) 
formulated in terms of the rescaled pressure function 
\be\label{Eq:stability-condition-resc}
	\eps M^\prime(\lambda)|_{\lambda=1} \, = \,
	\frac{\eps^2D}{B}\,\int\di^Dx\;\tilde{p}(x) = 0\, .
\ee

The expressions for $\eps M(\lambda)$ and $\eps M$ contain 
terms with explicit powers of $\teps^0$, $\teps^2$, $\teps^4$. 
As $\teps$ decreases the term $\propto\teps^0$ dominates,
and becomes the sole contribution to $\eps M$ in the limit 
$\eps\to 0$, where $\omega\to\omega_{\rm max}=m$ with the mass 
$m$ (\ref{Eq:mass-elementary-quantum}) 
of the elementary field $\Phi(x)$, such that we find 
\be\label{Eq:M-vs-Q-in-limit}
	\lim\limits_{\eps\to 0}\,\frac{M}{Q} = m,\,
\ee
meaning that the mass of the critical solution is entirely fixed 
in terms of its charge. This result can be deduced equivalently 
from (\ref{Eq:resc-densities},~\ref{Eq:resc-densities-def})
or (\ref{Eq:scaling-Esurf},~\ref{Eq:limit-M-vs-mQ}).

The virial theorem explains why the mass of the solution is 
fixed in terms of its charge: the term $\propto\eps^0$ in
(\ref{Eq:M-lambda}) is the leading contribution in $\eps M(\lambda)$, 
but drops out exactly from $\eps M^\prime(\lambda)$ at $\lambda =1$. 
The question of how the solution acquires stability  
(in the sense of a local minimum of the action) is therefore
answered by higher order terms in (\ref{Eq:M-lambda}) which contain 
information on the dynamics of the theory (\ref{Eq:Lagrangian}).
Thus, although they do not contribute to the mass of the critical
solution, the subleading terms in  (\ref{Eq:M-lambda}), which encode 
the dynamics of the theory, determine the shape of the solution.

It is instructive to clarify also the relative importance of the subleading 
terms proportional to $\teps^2$ and $\teps^4$ in (\ref{Eq:M-lambda}).
The variational problem of minimizing $\eps M(\lambda)$
receives information on the details of the theory, encoded 
in the dimensionless parameter $\alpha=2AC/B^2$, only through the 
subsubleading term $\propto\teps^4$ in (\ref{Eq:M-lambda}).

In the limit, the terms $\propto\teps^2$
determine the solution, while the term $\propto\teps^4$ 
becomes irrelevant and drops out. 
Thus, the critical solution depends on the
parameters $A$, $B$ of the theory (\ref{Eq:Lagrangian}),
but not on $C$. Moreover, the dependence on $A$, $B$ is
trivial in the sense that it provides trivial overall
prefactors. This is evident also from the equation of motion
which for $\eps\to0$ is given by
\be
	\rphi^{\prime\prime}(x)+\frac2x\;\rphi^\prime(x)-\rphi(x) +4\,
	\rphi^3(x) = 0 \, ,
	\label{Eq:eom-resc-limit} 
\ee
with boundary conditions as specified in (\ref{Eq:eom-resc}). 
This is a universal equation independent of the details of the 
theory (\ref{Eq:Lagrangian}) as encoded in the numerical values 
of the parameters $A$, $B$, $C$.
Equation (\ref{Eq:eom-resc-limit}) is actually universal for all 
(complex) $|\Phi|^4$ theories with a {\sl negative} coupling.
(Notice that it was crucial to include the factor $B^{-1/2}$ 
in the rescaling of the fields (\ref{Eq:rescaling}) in order to obtain 
the ``universal'' equation of motion (\ref{Eq:eom-resc-limit}) in 
the limiting case.)

We use the term {\sl universality} in this context to stress that
instead of $C|\Phi|^6$ in the potential (\ref{Eq:potential})
we could have had started with any positive $C^\prime|\Phi|^n$ term with even 
$n \ge 8$. After the rescaling and limiting procedure, we would 
have arrived at the same universal equation (\ref{Eq:eom-resc-limit}) 
with no memory of powers beyond $|\Phi|^4$ in the potential 
(\ref{Eq:potential}).

At this point it is important to realize that, while in the limiting 
case there is no memory of it in the rescaled equation of motion, 
the power beyond $|\Phi|^4$  is crucial for providing proper 
boundary conditions for the theory (\ref{Eq:Lagrangian}). In fact, 
the potential (\ref{Eq:potential}) would be unbound from below
without a positive term $|\Phi|^n$ with a power $n=6$ or higher.
Thus, although the explicit dependence on the parameter $C$ drops out,
it is crucial that the critical solution is understood as a careful
limiting procedure with a positive higher order term.

With the term $\propto C|\Phi|^6$ becoming irrelevant for $\eps\to 0$, 
the renormalizability of the theory (\ref{Eq:Lagrangian}) may seem
restored. But this is a subtle issue for two reasons. First, 
we deal with classical soliton solutions and quantum corrections 
(whose consideration is beyond the scope of this work) have to be 
considered \cite{Graham:2001hr}.
Second, a $|\Phi|^4$ theory with a negative coupling constant is 
actually ill-defined, as the potential is not bound from below. 
Thus, the limiting solution has to be understood 
within a careful limiting procedure with a however
small non-renormalizable term $\propto C|\Phi|^n>0$ 
in the potential with even $n =6$ or higher.

The limiting solution in the $Q$-ball system is reminiscent 
of the critical \cite{Bogomolny:1975de,Prasad:1975kr} 
monopole solution \cite{'tHooft:1974qc,Polyakov:1974ek} in 
the Georgi-Glashow model \cite{Georgi:1972cj,Shifman:2012zz} 
in three characteristic respects. 
First, the mass of the critical solution is fixed in terms of its 
charge, a property arising from symmetries of the Lagrangian 
(\ref{Eq:Lagrangian}) rather than its dynamics (which, of course, 
determines the shape of the limiting solution).
Second, it requires the presence of higher order terms in the original
Lagrangian which become irrelevant in the limit and whose only role
consists of providing boundary conditions for the theory.
Third, from (\ref{Eq:scaling-Esurf}) we obtain the inequality
$M\ge \omega Q$ which holds for all solutions, and becomes
saturated in the limit $\omega\to\omega_{\rm max}=m$. This imitates
the saturation of a Bogomol'nyi-type bound for critical monopole
solutions. Due to the simple U(1) symmetry, here
the bound takes a somewhat simplistic form.

The observation that $Q$-clouds correspond to universal 
non-perturbative solutions in a complex $|\Phi|^4$ theory with 
negative coupling may have interesting implications.
As the potential is not bound from below, taken by itself 
a $|\Phi|^4$ theory with a negative coupling is unphysical 
(unless supplemented by a however small positive, 
non-renormalizable higher order term, as discussed above). 
At the same time, it is connected by analytical continuation to a complex
$|\Phi|^4$ theory with a positive coupling. 
Analytical continuation is at the heart of the proof that in general 
perturbative expansions have zero convergence radius \cite{Dyson:1952tj}. 
A meaningful treatment of a theory is provided in the framework of 
``resurgent trans-series analysis'' \cite{Dunne:2012ae} where the 
perturbative series is combined with a series over all non-perturbative 
contributions, with the terms in both series connected to each other 
via specific ``resurgence relations.''
A question emerging in this context is whether our universal 
non-perturbative solution in complex $|\Phi|^4$ theory 
(with negative coupling constant) could contribute to
such a trans-series. This interesting question,
which is beyond the scope of our study, could potentially shed 
new light on the non-perturbative sector of complex $|\Phi|^4$ theories.

Finally, let us remark that irrespective of its interpretation, 
the limiting solution constitutes a local extremum of the action.
This is reflected by the fact that it satisfies the von Laue
condition, which means that the internal forces balance each
other exactly. This balance of forces is a necessary condition
for stability, but not sufficient. In fact, the $Q$-cloud solution 
constitutes a highly unstable field configuration: the smallest 
disturbances would result in formation of small and stable $Q$-balls.
This could make $Q$-clouds of interest as (toy) models for the 
inflationary era in the early universe, in particular if one 
succeeded in driving the limit $\eps\to0$ dynamically
\cite{Enqvist:2003gh}.

Worth mentioning in this context is the interesting connection of 
the $D$-term to the cosmological constant and inflation discussed in 
\cite{Teryaev:2013qba}.

Although in line with results from all other theoretical systems,
the negative $D$-term of the $Q$-cloud is still remarkable.
If even such an extremely unstable system has a negative $D$-term,
one may doubt whether it is possible to encounter a consistent system 
with a positive $D$-term. In this way our results contribute 
to the emerging understanding that $D$-terms are negative, 
which is rooted in the equilibrium of internal forces, 
even if it is a highly unstable equilibrium.

\section{Conclusions}
\label{Sec-8:conclusions}

We have presented a study of soliton solutions in the $Q$-ball 
system \cite{Coleman:1985ki} focusing on the limit where the frequency 
$\omega$ approaches its maximal value $\omega_{\rm max}$.

This limit was studied previously in \cite{Alford:1987vs} where it was
interpreted as the dissociation of unstable solutions into a dilute
gas of free quanta, and in \cite{Mai:2012yc} where properties
of the solutions were investigated and numerically found to exhibit 
characteristic scaling behavior, for example, mass $M\propto 1/\eps$ and 
$D$-term $\propto 1/\eps^2$ with $\eps = \sqrt{\omega^2_{\rm max}-\omega^2}$.
In this work, by working with adequately rescaled fields and coordinates,
we were able to go far beyond what was numerically tractable in 
\cite{Mai:2012yc}, and presented exact numerical solutions down to 
$\eps \ge 10^{-6}$ confirming qualitative findings of \cite{Mai:2012yc}. 
We studied in detail the solution of the rescaled equations 
of motion in the $Q$-cloud limit $\eps\to 0$ \cite{Alford:1987vs}, 
and showed how smoothly this limit is approached. 
All properties of the limiting solution exhibit, after appropriate scaling,
a smooth behavior as $\eps\to 0$. 
We derived analytical results for rescaled  quantities like 
$\eps\,M$ expressed in terms of the limiting solution.

The limiting solution has fascinating properties. If we include 
in the rescaling the parameter $B$ of the potential
$V=A\,\phi^2-B\,\phi^4+C\,\phi^6$, such that 
$\rphi(x)=\phi(r)/(\eps\,\sqrt{B})$ with $x=\eps\,r$,
we obtain in the limiting case a ``universal'' soliton equation 
of motion for the rescaled field $\rphi(x)$ which is independent
of the parameters $A$, $B$, $C$. More precisely, the parameters
$A$, $B$ provide trivial prefactors for the properties, 
while $C$ drops out in the limit, i.e.\ 
the sixtic term in the potential $V(\phi)$ becomes ``irrelevant''
in the sense of critical phenomena.

We showed that the limiting solution shares features of critical 
monopoles, and observed that it corresponds to a universal 
non-perturbative solution in complex $|\Phi|^4$ theory with negative 
coupling, which may have implications for the non-perturbative sector 
of $|\Phi|^4$ theories. 
But the main feature of the limiting solution is that it is a 
highly unstable field configuration. The smallest disturbance would 
cause a decay  into energetically favorable, small-size, stable 
$Q$-ball configurations. In this respect the $Q$-cloud resembles an 
undercooled gas, and could be of potential interest for (toy) 
models of the early universe.

Our initial motivation to study $Q$-clouds was triggered by the 
question whether such an extreme system with a genuine instability,
could exhibit a positive $D$-term. But despite the extreme instability,
the $Q$-cloud satisfies the von Laue condition. This means the internal
forces balance each other exactly, although one deals with a highly
unstable equilibrium situation. The balance of internal forces 
implies a negative sign for the $D$-term. Our study has
shown that conclusions regarding the sign of $d_1$ of
regular $Q$-balls \cite{Mai:2012yc} hold also for $Q$-clouds 
when viewed in terms of adequately scaled fields and coordinates. 
Considering the singularities associated with the $Q$-cloud limit, 
this result could not have been anticipated and required a careful
and dedicated study.

As the $Q$-cloud constitutes the most unstable system 
we are aware of, the finding of a negative $D$-term in this
extreme system is remarkable.
Our work does not prove that all $D$-terms are negative.
But considering that even such an extreme and unstable system
as a $Q$-cloud has a negative $D$-term, it is difficult to imagine
a consistent physical system with a positive $D$-term.

Our results support the emerging understanding that $D$-terms of 
particles are negative. 
It will be exciting to see whether this theoretical prediction
will be confirmed for nucleons and atomic nuclei in experiments 
at Jefferson Lab, CERN or the future Electron-Ion Collider 
\cite{Boer:2011fh}.

\

\

\noindent{\bf Acknowledgments.}
We thank Maxim Polyakov for discussions.
The work was partly supported by the National Science Foundation 
under Contract No.~1406298.

\

\appendix

\section{Systematic notation}
\label{App:systematic-notation}

The notation introduced in 
Eqs.~(\ref{Eq:resc-densities},~\ref{Eq:resc-densities-def}) was
very convenient for the discussion of the scaling behavior of the 
densities in Sec.~\ref{Sec-5:densities}. A notation showing more 
systematically the behavior of the charge density 
$\omega\,\rho_{\rm ch}(r)$ (where we include the factor of $\omega$
for convenience) and the EMT densities $T_{00}(r)$, $p(r)$, $s(r)$
is as follows
\begin{subequations}\label{Eq:lucid-rescaling}
\begin{alignat}
  \omega\omega\rho_{\rm ch}(r)\;=\;
	& 	\frac{(2A)^2}{B}\Biggl[ 
 		\teps^2\,\rphi(x)^2 
	&\;-\;&	\teps^4\,\rphi(x)^2  
        & &     &\Biggr]&,\label{Eq:lucid-rescaling-rho}\\
  T_{00}(r)\;=\;
	& 	\frac{(2A)^2}{B}\Biggl[	
       		\teps^2\,\rphi(x)^2 
     	&\;+\;& \teps^4\,\biggl\{\phantom{-}\frac12\,\rphi^\prime(x)^2 
		- \frac12\,\rphi(x)^2-\rphi(x)^4\biggr\}	
   	&\;+\;& 	\teps^6\,\alpha \,\rphi(x)^6
		&\Biggr]&,\label{Eq:lucid-rescaling-T00}\\
   p(r) 	\;=\;
	& 	\frac{(2A)^2}{B}\Biggl[	
     	&&	\teps^4\,\biggl(
		- \frac16\,\rphi^\prime(x)^2 
		- \frac12\,\rphi(x)^2
		+ \rphi(x)^4\biggr)
   	&\;-\;&	\teps^6\,\alpha \,\rphi(x)^6 
            	&\Biggr]&,\label{Eq:lucid-rescaling-p}\\
   s(r)	\;=\;	
     	&	\frac{(2A)^2}{B}\Biggl[	      
     	&& 	\teps^4\;\rphi^\prime(x)^2 
     	&&	&\Biggr]&.\label{Eq:lucid-rescaling-s}
\end{alignat}
\end{subequations}
This shows that $\omega\,\rho_{\rm ch}(r)$,
$T_{00}(r)$, $p(r)$, $s(r)$ contain only even powers of $\teps^n$ 
with $n=2,\,4,\,6$ with $\teps^2 = \eps^2/(2A)$ and the rescaled
fields and coordinates as defined in (\ref{Eq:rescaling}).
The curly brackets in Eq.~(\ref{Eq:lucid-rescaling-T00}) highlight 
the contribution $\propto\teps^4$ in $T_{00}(r)$ for later purposes.

The power $n=2$ is responsible for the leading (and in the limit sole) 
contributions to charge and energy densities $\omega\,\rho_{\rm ch}(r)$ 
and $T_{00}(r)$.
The power $n=4$ provides the dominant (and in the limit sole) terms in
the distributions of internal forces $s(r)$ and $p(r)$.
The power $n=6$ accompanies the ``irrelevant'' (from the point of view
of the limit) sixtic term from the potential, whose role is to provide
proper boundary conditions for the (non-renormalizable) theory,
as discussed in Sec.~\ref{Sec-7:interpretation}.

We remark that throughout this work we use $\eps$ or $\teps$
as defined in Eqs.~(\ref{Eq:epsilon},~\ref{Eq:teps}), 
depending which one yields a more convenient notation. 
At the expense of notational simplicity, we could have
worked with $\teps$ alone and preserved the dimensionality 
of fields and coordinates.

\section{\boldmath $M$ vs $mQ$ in the $Q$-cloud limit}
\label{App:limit-M-vs-mQ}

In this Appendix we show how $M$ approaches $mQ$,
and derive the result in (\ref{Eq:limit-M-vs-mQ}). This is of
interest, as exact analytic results are rare in soliton models.
We first prove that the expression highlighted by curly 
brackets in (\ref{Eq:lucid-rescaling-T00}) integrates to
a higher order term in $\teps$. 

Working with rescaled coordinates and fields, we define 
the ``equations of motion,'' ${\rm eom}(x)=0$, cf.\ 
Eq.~(\ref{Eq:eom-resc}), as
\ba
	{\rm eom}(x)=\rphi^{\prime\prime}(x)+\frac2x\rphi^\prime(x)
	-\rphi(x) +4\rphi^3(x) - 6\alpha\,\teps^2\rphi^5(x).
	\nonumber 
\ea
In this notation one can rewrite the term in the curly brackets in
(\ref{Eq:lucid-rescaling-T00}) as follows
\begin{align}\label{Eq-App:identity-local}
  x^2
  \biggl\{\,\frac12\,\rphi^\prime(x)^2 - \frac12\,\rphi(x)^2
    	    -\rphi(x)^4\biggr\} = - x^2{\rm eom}(x)\,\rphi(x) \;\;\;\;& 
  \nonumber\\
  + 3x^2\tilde{p}(x) 
    + \frac{\di}{\di x}\,\biggl[x^2\rphi(x)\rphi^\prime(x)\biggr]
    - 3\,\alpha\,\teps^2\,x^2\rphi(x)^6. &
\end{align}
The term proportional to ${\rm eom}(x)$ is zero anyway. Upon integration over 
$x$ the contribution of the pressure drops out due to the von Laue condition, 
Eqs.~(\ref{Eq:stability-condition},~\ref{Eq:stability-condition-resc}),
and so does the total derivative term. Thus, we find the identity
\be\label{Eq-App:identity}
	\int\di^3x\biggl\{\frac12\,\rphi^\prime(x)^2 - \frac12\,\rphi(x)^2
	    -\rphi(x)^4\biggr\} =
	-3\,\alpha\,\teps^2\!\int\di^3x\,\rphi(x)^6
\ee
which connects the subleading term in (\ref{Eq:lucid-rescaling-T00}) 
to the subsubleading term. In the limiting case $\teps\to 0$,
the right-hand-side of (\ref{Eq-App:identity}) yields exactly zero.
In particular, we obtain for the mass
\be\label{Eq-App:eps-M}
   \eps M =     \frac{2A}{B}\;\int\di^3x\;\biggl(\rphi(x)^2 
	        -2\,\alpha\,\teps^4\rphi(x)^6\biggr)\,.
\ee
The analog expression for $m\,Q$ is given by
\be\label{Eq-App:eps-mQ}
   \eps (m\,Q) =     \frac{2A}{B}\;\sqrt{1-\teps^2}\int\di^3x\;\rphi(x)^2 .
\ee
In this way we obtain the exact result
\be
	\frac{M-m\,Q}{M} 
	= \frac{1-\sqrt{1-\teps^2}-r\,\teps^4}{1-r\,\teps^4}\,,\;\;\;
	r = 2\,\alpha\,\frac{\int\di^3x\;\rphi(x)^6}{\int\di^3x\;\rphi(x)^2}\,,
\ee
and expanding in a series in $\teps$ we arrive at
\be\label{Eq-App:expansion-of-M-mQ}
	\frac{M-m\,Q}{M} = \frac12\,\teps^2 + \biggl(\frac18
	-r \biggr)\,\teps^4
	+{\cal O}(\teps^6)\,.
\ee
We see that $M$ approaches $m\,Q$ from above, and find that the first 
term in the small-$\teps$ expansion of the expression for $(M-m\,Q)/M$ 
can be evaluated analytically. Higher order terms in this expansion
in general cannot be computed exactly, which we have indicated
by quoting the exact coefficient of the $\teps^4$-term in 
(\ref{Eq-App:expansion-of-M-mQ}).

It is interesting to express Eq.~(\ref{Eq-App:eps-M}) in terms of the 
original fields and coordinates, which yields 
\be\label{Eq-App:M-new-relation}
   M = \int\di^3r\;\biggl(2A\,\phi(r)^2-2\,C\,\phi(r)^6\biggr)\,.
\ee
Thanks to the identity (\ref{Eq-App:identity-local}), the mass 
(of any solution) can be expressed in terms of two specific terms in
the potential (\ref{Eq:potential}).
We are not aware that this relation has been derived before in literature.
From (\ref{Eq-App:M-new-relation}) we see that
always $M\le  m^2 \int\di^3r\;\phi(r)^2$, where the equal
sign holds only in the $Q$-cloud limit when the contribution
of $\,C\,\phi(r)^6$ becomes irrelevant.

\section*{References}


\end{document}